\documentclass[amsmath,pbl,
showpacs, prl, twocolumn] {revtex4}
\usepackage{bm}
\usepackage{enumerate}
\usepackage{graphicx}
\usepackage[dvips]{epsfig}
\usepackage{epsf}
\usepackage{xcolor}
\makeatletter
\newcommand{\beq}{\begin{equation}}
\newcommand{\eeq}{\end{equation}}
\newcommand{\bea}{\begin{eqnarray}}
\newcommand{\eea}{\end{eqnarray}}
\newcommand{\bse}{\begin{subequations}}
\newcommand{\ese}{\end{subequations}}
\newcommand{\bwt}{\begin{widetext}}
\newcommand{\ewt}{\end{widetext}}
\newcommand{\Rmnum}[1]{\expandafter\@slowromancap\romannumeral #1@}

\makeatletter

\begin{document}
\title{Dirac magnons pairing via pumping}
\author{Vladimir A. Zyuzin}
\affiliation{Nordita, KTH Royal Institute of Technology and Stockholm University, Roslagstullsbacken 23, SE-106 91 Stockholm, Sweden}
\begin{abstract}

We study pumping of magnons to the Dirac points of magnon's Brillouin zone of a ferromagnet on a honeycomb lattice. 
In particular, we consider second-order Suhl process, when due to interaction between magnons, a pair of magnons is created due to absorption of two electromagnetic wave quanta. 
We introduce a bosonic analog of the Cooper ladder for the magnon pair, which is shown to enhance the pairing of magnons at the Dirac points.  
As a result of pairing of the Dirac magnons, the system becomes unstable towards formation of a magnetic state with zero or reduced magnetization - the Dirac magnon paired state. 
In this case the resonant frequency of the pump equals to that of energy of the Dirac points. 
Our estimates suggest that the Dirac magnon paired state can be found in the CrBr$_{3}$ or CrCl$_{3}$ ferromagnet below in vicinity of the Curie temperature.
\end{abstract}
\maketitle

Magnons are fluctuations about the spontaneous magnetic order. 
Typically two types of magnons are distinguished based on the magnetic structure, ferromagnetic or antiferromagnetic. The two have different low-energy, low-momentum dispersion, regardless of the lattice structure of the magnetic structure. Ferromagnetic magnons are quadratic in momentum, while antiferromagnetic are linear (for example, see \cite{ABP1967,Auerbach,Rezende}). Recently, because of the progress made in topological properties of fermions, a topology tool has been applied to understand intrinsic transport properties of magnons. With that details of the lattice structure became important. 
Certain lattices, for example, pyrochlore \cite{Onose2010, ZhangPRB2013}, kagome \cite{ZhangPRB2013,Katsura2010,MookHenkMertig, HirschbergerPRL2015, KovalevZyuzin}, and honeycomb \cite{MaksimovChernyshev, Fransson, Owerre2016a, Owerre2016b, Kim2016, KovalevZyuzinLi, Pershoguba, YelonSilberglitt,HoneycombPRX2018,HoneycombPRX2020}, allow for natural magnons's momentum-pseudospin locking. 
In ferromagnets such locking results in Dirac crossing points (degeneracies) at some particular high-energy and finite-momentum points in magnon's Brillouin zone. 
It is convenient to call magnons at such crossing points as the Dirac magnons \cite{Fransson}. 
As a result of the locking, certain types of the Dzyaloshinskii-Moriya interaction, allowed by the lattice symmetry, result in various transverse responses of magnons to the temperature gradient, such as magnon thermal Hall \cite{Onose2010, Katsura2010, Owerre2016b} and Nernst effects \cite{KovalevZyuzin}, and to fictituous gauge fields \cite{KovalevZyuzinLi}, such as the magnon Hall effect.

\begin{figure}[t] 
\centerline{
\begin{tabular}{ccc}
\includegraphics[width=0.24 \columnwidth]{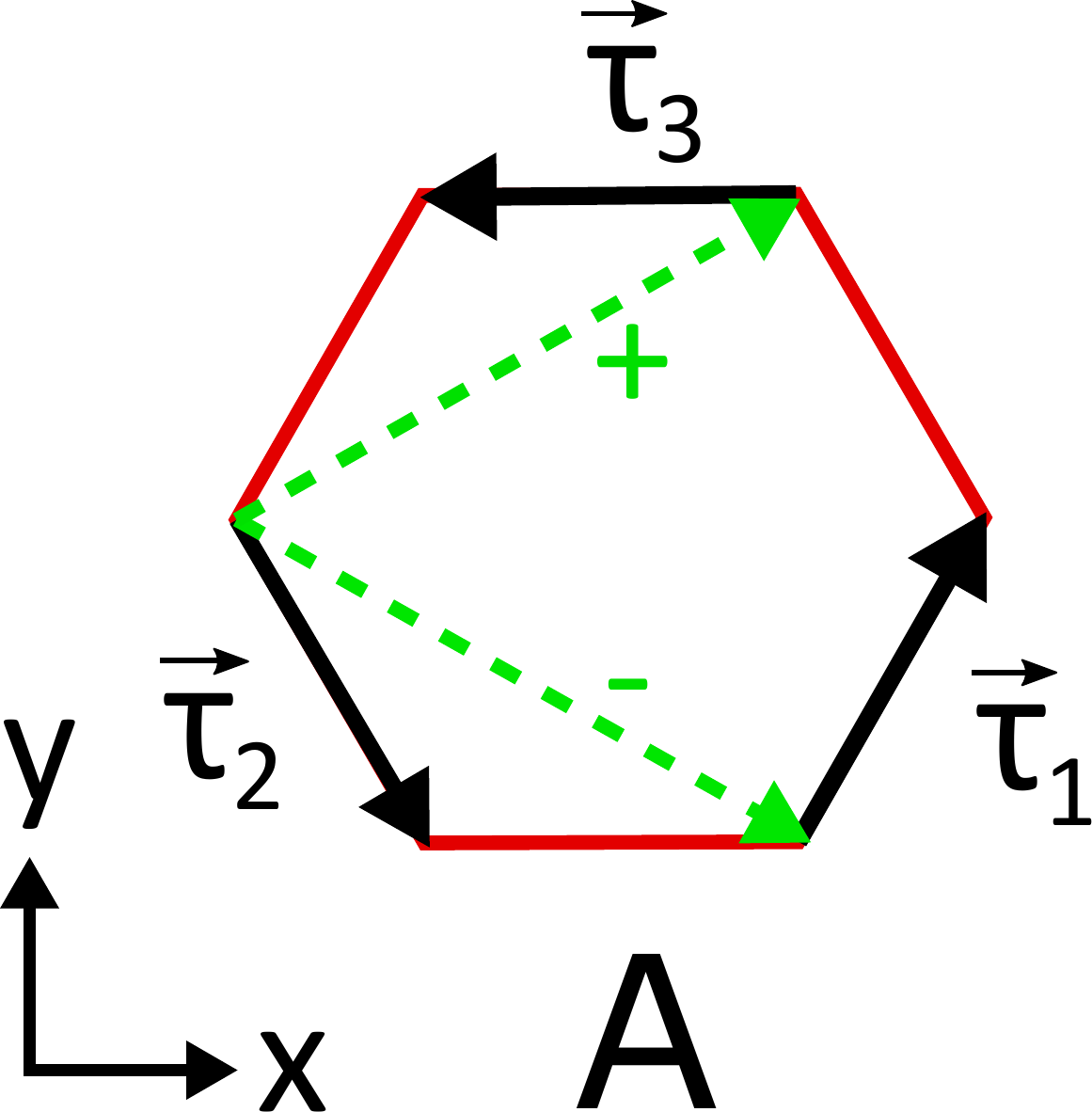}&
\includegraphics[width=0.45 \columnwidth]{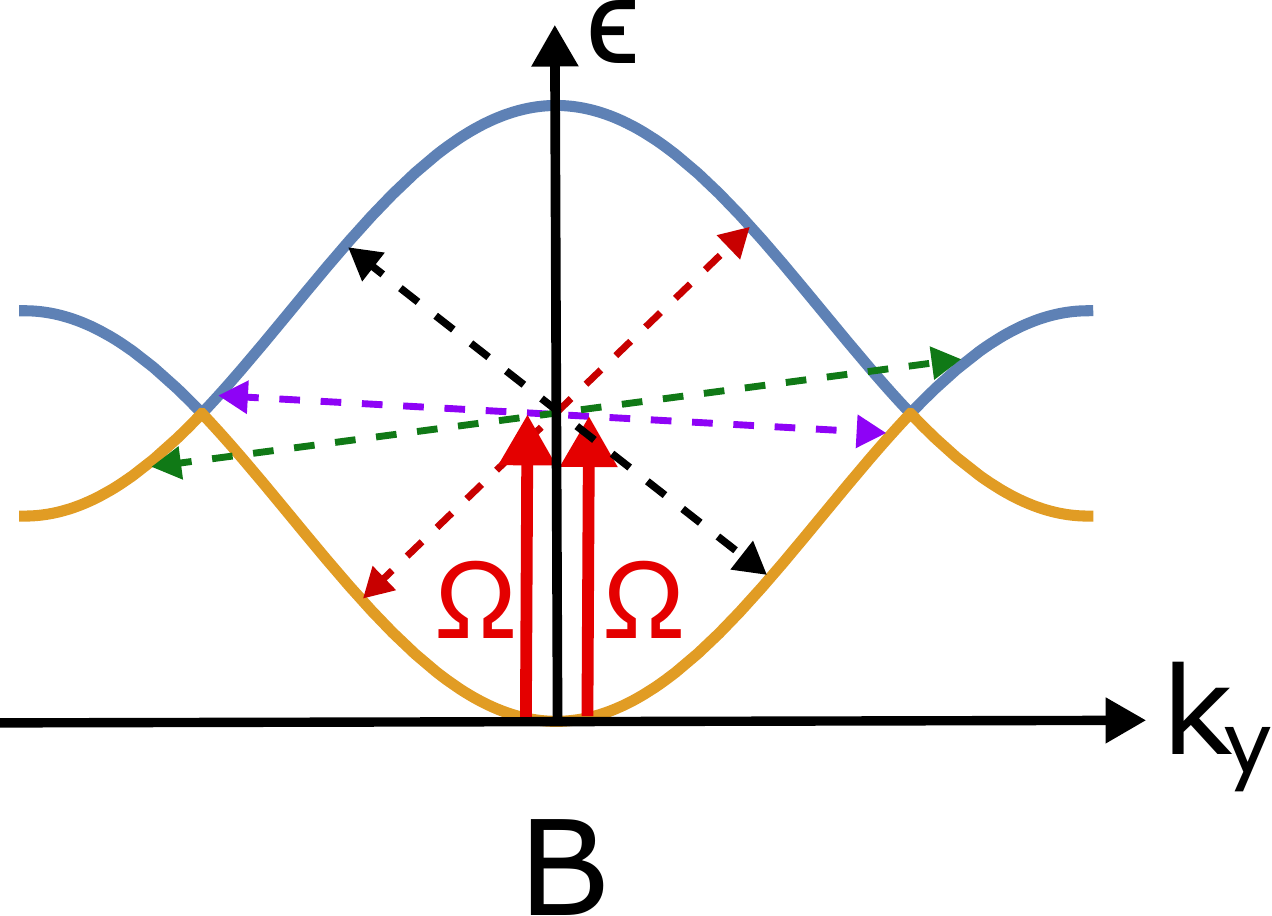}&
\includegraphics[width=0.24  \columnwidth]{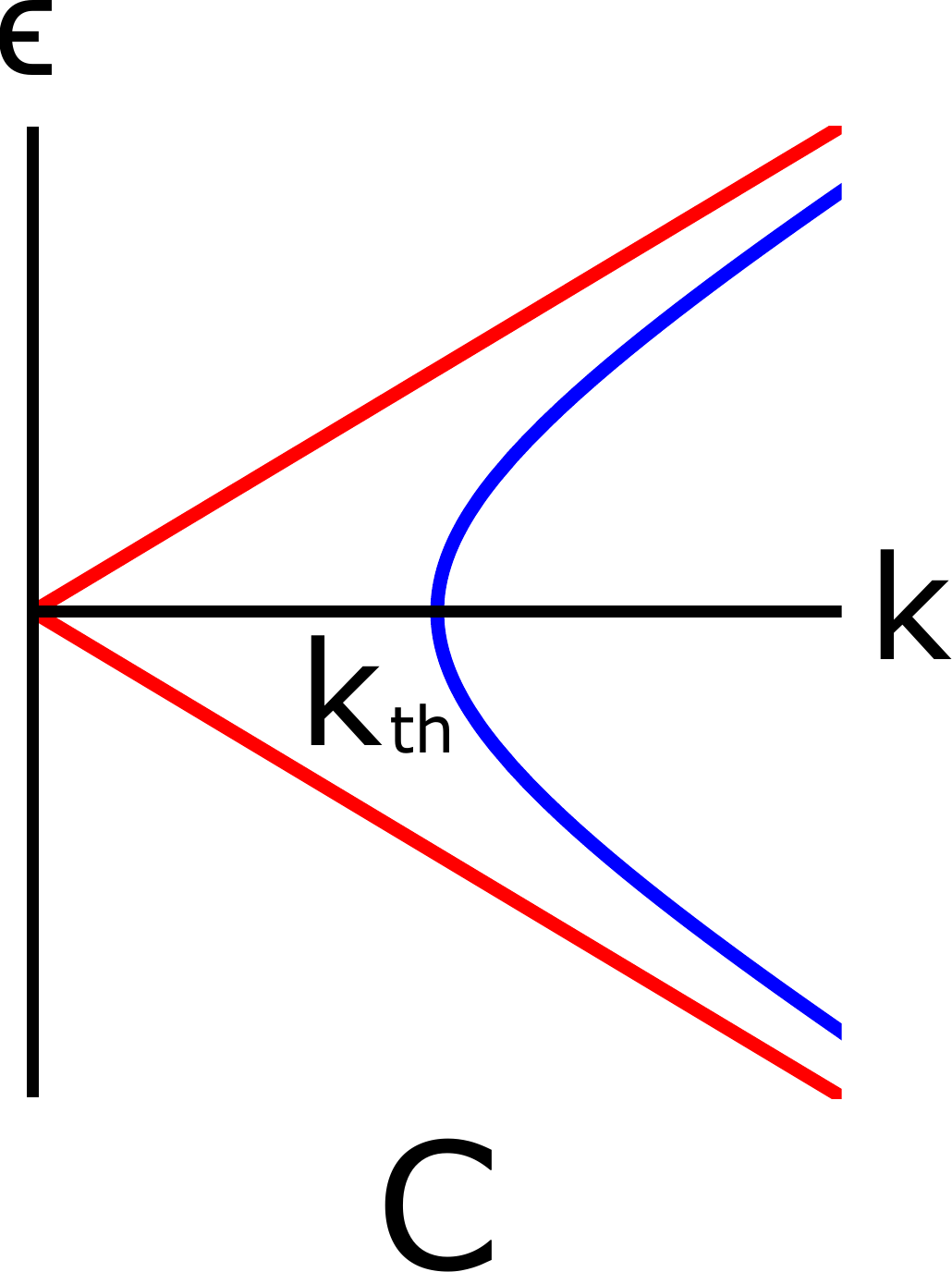}\\
\end{tabular}
}
\protect\caption{A. Schematics of the honeycomb lattice. Vectors connecting the nearest neighbor cites are ${\bm \tau}_{1}=\frac{1}{2}\left( \frac{1}{\sqrt{3}},1\right)$, ${\bm \tau}_{2} = \frac{1}{2}\left( \frac{1}{\sqrt{3}},-1\right)$, and ${\bm \tau}_{3} = \frac{1}{\sqrt{3}}(-1,0)$.
B. $k_{x} =0$ slice of spectrum of free magnons. Dashed lines describe examples of different pairs with ${\bf k}$, $\Omega +\epsilon$ and $-{\bf k}$, $\Omega -\epsilon$ momenta and frequency, for pump's frequency $\Omega = 3SJ$. C. Magnon spectra below in vicinity of the Dirac point: original (red) linear spectrum of free magnons and spectrum of magnon pairs (blue) with a threshold $k_{\mathrm{th}}$ defined in Eq. (\ref{pair}).}

\label{fig:pair}  

\end{figure}

\begin{figure}[t] 
\centerline{
\includegraphics[width=0.9  \columnwidth]{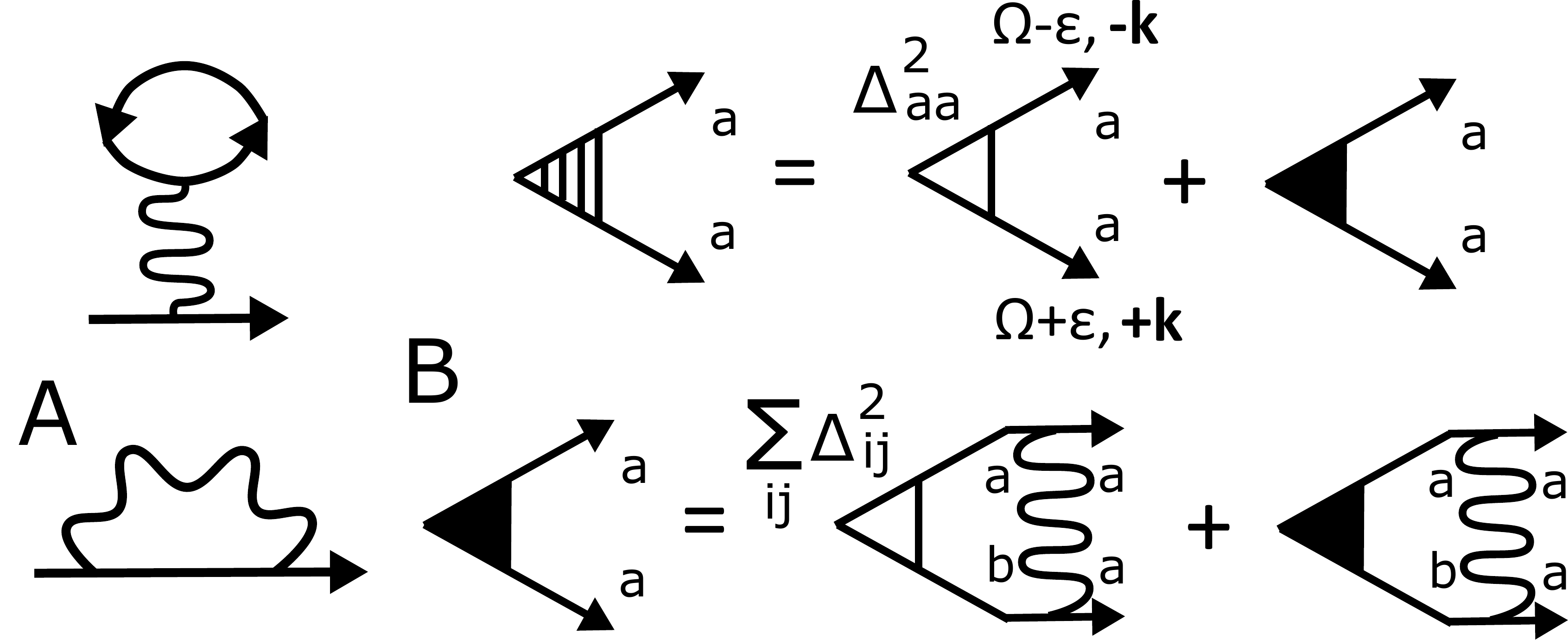}
}
\protect\caption{A. Hartree-Fock corrections to the dispersion of magnons. Wavy lines stand for the interaction defined in Eq. (\ref{interaction}). B. Graphic equation for the pairing interaction strength. Here empty triangle stands for the initial pairing interaction strength $\Delta^2_{ij}$ defined in accordance with Eq. (\ref{secular}), $\Delta^2_{\mathrm{aa}}=\Delta^2_{\mathrm{bb}} =- \Delta^2 \gamma_{0}$, and $\Delta^2_{\mathrm{ab}} =(\Delta^2_{\mathrm{ba}})^{*} = \Delta^2\gamma_{\bf k}$. 
Black tringle is intermediately renormalized pairing interaction strength, and lined triangle is the overall renormalized pairing interaction strength.}

\label{fig:vertex}  

\end{figure}

In this Letter we find another unique Dirac magnons property revealed under the second-order Suhl magnon pumping process \cite{Suhl1957,ZakharovLvovStarobinets,Rezende,ABP1967}. 
A single Dirac magnon can't be created in a process of absoroption of one pumping field quanta. 
This is because such magnon is located at non-zero momentum in the Brillouin zone, and there is no way to conserve the momentum in the process of absoroption, as the experimentally relevant pumping field has a zero wave vector. 
However, a pair of magnons with opposite momenta can be created when two pump field quanta are absorbed. 
Such processes are known as the second-order Suhl processes (see Fig. \ref{fig:pair}B). 
We show that this process is not present in linearized spin-wave theory of magnons, but appears when the interaction between the magnons is included to the consideration.
The frequency of the pump can scan the entire Brillouin zone of the magnons, and the absorption of two magnons can happen at any frequency.
However, as we show in this Letter, the pump's frequency equal to the energy of the Dirac points is the resonant due to the
magnon rescattering processes of the Cooper ladder type (see Fig. \ref{fig:vertex}B). 
This is because at such frequency, the system can accomodate the largest amount of magnon pairs with opposite momenta and frequencies (see Fig. \ref{fig:pair}B). 
For frequencies away from the Dirac points, the pairing of two magnons is parametrically weakened by the rescattering processes.
The resonance corresponds to an instability of the system towards formation of a zero magnetization state. 
Below, we refer to such resonance as the Dirac magnons paired state.
We hope Dirac magnons paired state can be experimentally observed in ferromagnets with spins on pyrochlore lattice \cite{Onose2010}, or layered kagome \cite{HirschbergerPRL2015} and honeycomb lattices \cite{YelonSilberglitt,HoneycombPRX2018,HoneycombPRX2020}. 
In particular, based on our estimates, we predict that it can be observed in the honeycomb CrBr$_{3}$ ferromagnet \cite{YelonSilberglitt} below in vicinity of the Curie temperature.

To demonstrate the effect, let us study a model of insulating ferromagnet in which spins of length $S$ are located on the sites of honeycomb lattice (see Fig. \ref{fig:pair}A). 
Near neighbor spins interact with each other via ferromagnetic Heisenberg exchange interaction. Ferromagnetic order is assumed to be in $z$-direction, this can be achieved by applying a small magnetic field in $z-$ direction. There is a pumping field which is perpendicular to the order, and which oscillates with a frequency $\Omega$ and has a zero wave vector. 
Hamiltonian of the system reads,
\begin{align}\label{model}
H
= -J\sum_{\langle ij \rangle}  {\bf S}_{i}{\bf S}_{j}  + \Gamma \sum_{i} \left[ S^{x}_{i}\cos(\Omega t) + S^{y}_{i}\sin(\Omega t) \right] ,
\end{align}
where $J>0$ is the exchange couping energy and $\Gamma$ is the pump's intensity. 
In order to study the spin-waves, we use the Holstein-Primakoff presentation of spin operators in terms of bosons, namely
for A atoms (there are two atoms in the unit cell, A and B) operators $S_{i}^{\pm} = S_{i}^{x}\pm i S_{i}^{y}$ and $S_{i}^{z}$ read as
$S_{i}^{+} = \sqrt{2S - a_{i}^{\dag}a_{i}}a_{i}$, $S_{i}^{-} = a_{i}^{\dag}\sqrt{2S - a_{i}^{\dag}a_{i}}$, and $S_{i}^{z} = S - a_{i}^{\dag}a_{i}$, with $[a_{i},a_{j}^{\dag}]=\delta_{i,j}$ boson commutation relation. The same is performed for the B atoms with the help of $b_{i}$ and $b_{i}^{\dag}$ boson operators.

In the space of elements of the honeycomb's unit cell, in which case the boson operators are defined by $\Psi_{\bf k}^\dag = ( a_{\bf k}^{\dag}, ~b_{\bf k}^{\dag})$, the Hamiltonian of non-interacting spin-waves reads as
\begin{align}\label{linear}
H_{0} = 
SJ
\int_{\bf k}
\Psi_{\bf k}^\dag
\left[ 
\begin{array}{cc}
3 & - \gamma_{\bf k} \\
- \gamma_{\bf k}^{*} & 3
\end{array}
\right]
\Psi_{\bf k} 
\equiv 
\int_{\bf k}
\Psi_{\bf k}^\dag
[\hat{H}_{0}]_{\bf k}
\Psi_{\bf k},
\end{align}
where $\gamma_{\bf k}=\sum_{i=1,2,3}e^{i{\bf k}{\bm \tau}_{i}} = 2e^{i\frac{k_{x}}{2\sqrt{3}}}\cos\left( \frac{k_{y}}{2}\right)+ e^{-i\frac{k_{x}}{\sqrt{3}}}$ is the nearest-neighbor hopping element see left part of the Fig. \ref{fig:pair}, and $\int_{\bf k} \equiv \int\frac{d{\bf k}}{(2\pi)^2}$ for two-dimensional system.
Diagonalization gives energy spectrum of non-interacting magnons,
$\varepsilon_{\pm; {\bf k}} =SJ\left( 3 \pm  \vert \gamma_{\bf k}\vert \right)$
with corresponding wave functions
$\varphi_{\pm} = 
\frac{1}{\sqrt{2}}
[ 
\mp\frac{\gamma_{\bf k}}{\vert \gamma_{\bf k} \vert} ,~
1
]^{\mathrm{T}}$. 
At special ${\bf K}=(0,-\frac{4\pi}{3})$ and ${\bf K}^\prime =(0,\frac{4\pi}{3})$ points the spectrum is linear and is described by the Dirac Hamiltonian. The energy of magnons at these points is $\varepsilon_{\pm;{\bf k}} = 3SJ$.
Terms quartic in boson operators describe interactions between the magnons. Normal ordered interaction reads
\begin{align}\label{interaction}
&
H_{\mathrm{int}} =
-J\int_{\{ {\bf k}\}}\delta_{\{{\bf k}\}}  
\gamma_{{\bf k}_{4}-{\bf k}_{3}} 
a_{{\bf k}_{1}}^{\dag}b_{{\bf k}_{3}}^{\dag} a_{{\bf k}_{2}} b_{{\bf k}_{4}}
\\
&
+ \frac{J}{4}\int_{\{ {\bf k}\}}
\delta_{\{{\bf k}\}} 
\left[
\gamma_{{\bf k}_{3}}^{*}
 a_{{\bf k}_{1}}^{\dag}b_{{\bf k}_{3}}^{\dag} a_{{\bf k}_{2}}a_{{\bf k}_{4}}
+
\gamma_{{\bf k}_{3}}
 a_{{\bf k}_{2}}^{\dag}a_{{\bf k}_{4}}^{\dag} a_{{\bf k}_{1}}b_{{\bf k}_{3}}
\right]
\nonumber
\\
&
+ \frac{J}{4}\int_{\{ {\bf k}\}}
\delta_{\{{\bf k}\}} 
\left[
\gamma_{{\bf k}_{1}}
 a_{{\bf k}_{1}}^{\dag}b_{{\bf k}_{3}}^{\dag}  b_{{\bf k}_{2}}b_{{\bf k}_{4}}
+
\gamma_{{\bf k}_{1}}^{*}
 b_{{\bf k}_{2}}^{\dag}b_{{\bf k}_{4}}^{\dag}  a_{{\bf k}_{1}}b_{{\bf k}_{3}}
\right],
\nonumber
\end{align}
where $\delta_{\{{\bf k}\}} \equiv \delta_{ {\bf k}_{1}-{\bf k}_{2},{\bf k}_{4}-{\bf k}_{3}}$ short notation was used, and $\int_{\{{\bf k} \}}$ stands for integration over all momenta. The interaction is instanteneous in time.
Hamiltonian in momentum space describing pump field with a frequency $\Omega$ is
\begin{align}\label{pump}
H_{\mathrm{pump}} 
=
\frac{\Gamma\sqrt{S}}{\sqrt{2}}  \left[ (a_{0}+b_{0})e^{-i\Omega t} + ( a^{\dag}_{0}+b^{\dag}_{0})e^{i\Omega t}  \right],
\end{align}
where $a_{0} \equiv a_{{\bf k} = 0}$ and the same for $b_{0}$. 
In order to understand the effect of the pumping field Eq. (\ref{pump}) on the magnons described by Eq. (\ref{linear}) and (\ref{interaction}), we study the system in the Keldysh time space.
This space complicates the analysis but gives a clear understanding of all relevant processes. 
We promote boson fields $a_{\bf k},b_{\bf k}$ to $\Psi_{\alpha;{\bf k};\epsilon}, \Psi_{\beta;{\bf k};\epsilon}$ fields, in which frequency $\epsilon$ was explicitly used, and $a^{\dag}_{\bf k},b^{\dag}_{\bf k}$ to $\bar{\Psi}_{\alpha;{\bf k};\epsilon}, \bar{\Psi}_{\beta;{\bf k};\epsilon}$. Furthermore, the fields are promoted to classical ($\mathrm{cl}$) and quantum components ($\mathrm{q}$) components in accord with the Keldysh technique (see Supplemental Material \cite{SM} for details and, for example, \cite{Kamenev}).
Let us show how to conveniently capture the process of absorption of the pumping field Eq. (\ref{pump}) by the magnons.

We write the advanced part of the Lagrangian describing non-interacting magnons defined by Eq. (\ref{linear}) with $\epsilon = \Omega$ frequency and ${\bf k} = 0$ momentum,
\begin{align}
{\cal L}^{\mathrm{A}}_{0,\Omega} = \sum_{m,n}\bar{\Psi}^{\mathrm{cl}}_{m, 0, \Omega} \hat{{\cal L}}^{\mathrm{A}}_{mn,0,\Omega} \Psi^{\mathrm{q}}_{n, 0, \Omega}
-\Gamma\sqrt{S}\sum_{n} \Psi^{\mathrm{q}}_{n, 0, \Omega}  ,
\end{align}
where $\hat{{\cal L}}^{\mathrm{A}}_{mn,{\bf k},\Omega} = (\Omega -i0)\delta_{mn} - [\hat{H}_{0}]_{mn,{\bf k}}$ is the Lagrangian density desribing corresponding non-interacting magnons, and $m,n=\{\alpha, \beta\}$ are the indeces describing pseudospin.
We now want to get rid of the terms linear in $\Psi^{\mathrm{q}}_{n, 0, \Omega}$ in the action with the following shift,
\begin{align}\label{shift}
&
\bar{\Psi}^{\mathrm{cl}}_{\alpha, 0, \Omega} \rightarrow \bar{\Psi}^{\mathrm{cl}}_{\alpha, 0, \Omega} + x_{\mathrm{A}}, 
~~~~
\bar{\Psi}^{\mathrm{cl}}_{\beta, 0, \Omega} \rightarrow \bar{\Psi}^{\mathrm{cl}}_{\beta, 0, \Omega} + y_{\mathrm{A}}, 
\end{align}
where $x_{\mathrm{A}}$ and $y_{\mathrm{A}}$ are obtained to be
\begin{align}
&
x_{\mathrm{A}}= \frac{{\cal L}^{\mathrm{A}}_{\beta\alpha,0,\Omega} - {\cal L}^{\mathrm{A}}_{\beta\beta,0,\Omega} }{ {\cal L}^{\mathrm{A}}_{\alpha\beta,0,\Omega}{\cal L}^{\mathrm{A}}_{\beta\alpha,0,\Omega}   - {\cal L}^{\mathrm{A}}_{\beta\beta,0,\Omega}{\cal L}^{\mathrm{A}}_{\alpha\alpha,0,\Omega}}\Gamma\sqrt{S},
\\
&
y_{\mathrm{A}}= \frac{{\cal L}^{\mathrm{A}}_{\alpha\beta,0,\Omega} - {\cal L}^{\mathrm{A}}_{\alpha\alpha,0,\Omega} }{ {\cal L}^{\mathrm{A}}_{\alpha\beta,0,\Omega}{\cal L}^{\mathrm{A}}_{\beta\alpha,0,\Omega}   - {\cal L}^{\mathrm{A}}_{\beta\beta,0,\Omega}{\cal L}^{\mathrm{A}}_{\alpha\alpha,0,\Omega}}\Gamma\sqrt{S}.
\end{align}
The same procedure is performed for the retarded part of the action to take care of the $-\Gamma\sqrt{S}\sum_{n} \bar{\Psi}^{\mathrm{q}}_{n, 0, \Omega}$ linear term. See \cite{SM} for more details. After the shifts are performed, the non-interacting action is the same as the one without the linear terms. Even the Keldysh part of the action does not get affected. 
However, the shifts drastically modify terms describing magnon-magnon interactions Eq. (\ref{interaction}). In particular, new terms in the bilinear part of the Hamiltonian describing magnons with any frequency and momentum, rather than the pumped ones with $\Omega$ and ${\bf k}=0$, appear. 
Our calculations show (see \cite{SM} for details) that there is no way to obtain terms of the $\propto\bar{\Psi}^{\mathrm{cl}/\mathrm{q}}_{n,{\bf k},\epsilon}\Psi^{\mathrm{q}/\mathrm{cl}}_{m,{\bf k},\epsilon}$ type, but new terms describing pairing \cite{Suhl1957} of two magnons, i.e. of the $\propto\bar{\Psi}^{\mathrm{cl}/\mathrm{q}}_{n,{\bf k},\Omega+\epsilon}\bar{\Psi}^{\mathrm{q}/\mathrm{cl}}_{m,-{\bf k},\Omega-\epsilon}$ type, appear. 
Physically, they originate due to the absorbtion of two pump's quanta, and describe a subsequent creation of a magnon pair with ${\bf k},\Omega+\epsilon$ and $-{\bf k},\Omega-\epsilon$ momenta and energies (see Fig. \ref{fig:pair}B).

Let us now understand what will the creation of a magnon pair do to the system. Our calculations show that in the extended space of magnons, $\bar{\Phi}^{\mathrm{cl}/\mathrm{q}}_{{\bf k},\epsilon} =\frac{1}{\sqrt{2}}( \bar{\Psi}^{\mathrm{cl}/\mathrm{q}}_{\alpha,{\bf k},\Omega+\epsilon}, \bar{\Psi}^{\mathrm{cl}/\mathrm{q}}_{\beta,{\bf k},\Omega+\epsilon},\Psi^{\mathrm{cl}/\mathrm{q}}_{\alpha,-{\bf k},\Omega-\epsilon}, \Psi^{\mathrm{cl}/\mathrm{q}}_{\beta,-{\bf k},\Omega-\epsilon})$, the spectrum of a pair of magnons is given by a solution of the following secular equation,
\begin{align}\label{secular}
\mathrm{det}
&
\left[ 
\begin{array}{cccc}
\zeta+\epsilon  & SJ\gamma_{\bf k} & -\Delta^2\gamma_{0} & \Delta^2\gamma_{{\bf k}} \\
SJ\gamma_{\bf k}^{*} & \zeta+\epsilon   & \Delta^2\gamma^{*}_{{\bf k}} & -\Delta^2\gamma_{0} \\
-\Delta^2\gamma_{0} & \Delta^2\gamma_{{\bf k}} & \zeta - \epsilon  & SJ\gamma_{\bf k} \\
\Delta^2\gamma_{{\bf k}}^{*} & -\Delta^2\gamma_{0} & SJ\gamma_{\bf k}^{*} & \zeta - \epsilon 
\end{array}
\right] 
= 0,
\end{align}
where $\zeta = \Omega - 3SJ$ is introduced for brevity, $\gamma_{0}=3$, and where the pairing strength $\Delta^2 = \frac{J}{4}\left( \frac{\Gamma \sqrt{S}}{3SJ}\right)^2$ for $\zeta = 0$, and $\Delta^2 = \frac{J}{4}\left( \frac{\Gamma \sqrt{S}}{6SJ}\right)^2$ for $\zeta = \pm3SJ$ and otherwise according to the shift Eq. (\ref{shift}), was defined. 
The equation is the boson analog of the Bogoliubov-de Gennes Hamiltonian in fermion systems. 
The difference is in the structure of signs of the frequencies $\epsilon$ on the main diagonal in Eq. (\ref{secular}).
The spectrum reads
\begin{align}\label{pair}
\epsilon_{\pm;{\bf k}}^{2} =  (\zeta \pm SJ\vert \gamma_{\bf k} \vert)^2 - \Delta^4(\gamma_{0}  \mp \vert\gamma_{\bf k}\vert )^2.
\end{align}
Therefore, the system of pumped interacting magnons will become unstable when $\epsilon_{\pm; {\bf k}}^{2} < 0$ is satisfied. Let us analyze different parts of the spectrum for such an instability.

Let us first study a special case when pump's frequency is $\Omega = 3SJ$ for which $\zeta =0$.
Then, at the ${\bm \Gamma}=(0,0)$ point $\vert \gamma_{\bf k}\vert \approx 3 -\frac{k^2}{4}$, then $\epsilon_{+;{\bf k}}^{2} =  (SJ)^2\left(3-\frac{k^2}{4} \right)^2 - 36\Delta^4$. For the instability to occur at the ${\bm \Gamma}$ point, the intensity of the pump should become larger than the exchange coupling energy. However, experimentally reasonable assumption is $SJ > \Delta^2$ which means it is impossible to make the system unstable at the ${\bm \Gamma}$ point.
On the other hand, at the ${\bf K}$ and ${\bf K}^\prime$ we approximate $\vert \gamma_{\bf k} \vert \approx \frac{\sqrt{3}}{2}k$, and get for the spectrum $\epsilon_{\pm;{\bf k}}^{2} \approx  (SJ)^2 \frac{3}{4}k^2 - 9\Delta^4$. From here we observe that the solution is always unstable for momenta smaller than the threshold value of $k_{\mathrm{th}} = \frac{2\sqrt{3}\Delta^2}{SJ}$, i.e. for $k<k_{\mathrm{th}}$. For schematics see Fig. \ref{fig:pair}C.

Having pumped the magnons to the Dirac points, let us now study their rescattering processes. 
In the first order in interaction Eq. (\ref{interaction}) we get Hartree-Fock type corrections shown in Fig. \ref{fig:vertex}A to the magnon's dispersion \cite{BlochPRL1962}. See SM for the details of their derivation.
Interaction Eq. (\ref{interaction}) treated to second order contributes to the magnon's life-time \cite{Pershoguba}. 
Here we study how the pairing interaction strength $\Delta^2$ gets renormalized by the interaction. 
For that we contstruct a boson analog of the Cooper ladder shown in Fig. \ref{fig:vertex}B.
Our calculations show (see \cite{SM} for details) that the operator structure of $\Delta^2$ given in Eq. (\ref{secular}) gets reproduced at each step of the ladder. Then, summing up the ladder, we replace $\Delta^2$ for $\zeta=0$ with
\begin{align}\label{ladder}
\Delta^2 \rightarrow
\frac{\Delta^2}{ 1-  \frac{J}{\tilde{J}}  \left[\frac{1}{4S} + \frac{\pi }{8S}\left(\frac{T}{3S\tilde{J}}\right)^2\right] },
\end{align}
where $\tilde{J} = J\left[1-\frac{\pi}{4S}\left( \frac{T}{3SJ}\right)^2 \right]$ includes the Hartree-Fock corrections.
The integral defining a step of the ladder is counting the number of pairs which can be created for a given frequency.
Clearly the pairing of Dirac magnons is enhanced due to the rescattering processes. 
The minus sign in the denominator in Eq. (\ref{ladder}) is due to the repulsive nature of the last two terms in Eq. (\ref{interaction}). 
Our estimates suggest that for honeycomb lattice CrBr$_{3}$ or CrCl$_{3}$ $S=\frac{3}{2}$ ferromagnet \cite{YelonSilberglitt}, the tendency is such that at temperatures below in vicinity of the Curie temperature, i.e. $T \simeq T_{c} \sim 3SJ$, the expression for the renormalized pairing strength Eq. (\ref{ladder}) diverges. 
This signals a transition to a new state, which we call Dirac magnons paired state. 
It seems natural that the transition occurs below in vicinity of the Curie temperature, as there are plenty of magnons in the system and their rescattering processes are known to become important \cite{Pershoguba, BlochPRL1962,YelonSilberglitt}. If the spin is made more classical by increasing its length $S$, the denominator in Eq. (\ref{ladder}) does not become singular for any temperature.

Let us study the effect of Dzyaloshinskii-Moriya interaction of the $
H_{\mathrm{DMI}} = D\sum_{\langle \langle ij \rangle\rangle}\nu_{ij}[{\bf S}_{i}\times {\bf S}_{j}]_{z}
$ type on the pairing. Here $D$ is a constant, $\langle \langle ij \rangle\rangle$ notation counts second-nearest neighbors, and $\nu_{ij} = \pm 1$ is defined by the green dashed arrows in Fig. \ref{fig:pair}A (see SM for more details).
In the vicinity of the Dirac points, i.e $\zeta = 0$, the spectrum of magnon pairs is now 
\begin{align}
\epsilon^2_{\pm} = (SJ)^2\frac{3}{4}k^2 + \chi^2  - 9\Delta^4,
\end{align}
where $\chi =3\sqrt{3}SD$.
We conclude that if $\vert \chi \vert  \geq  3\Delta^2$ there will be no instability in the system. In unpumped ferromagnet such Dzyaloshinskii-Moriya interaction opens up a gap at the Dirac points in the spectrum of the magnons. Then, for the Dirac magnons paired state to occur, pumping strength should overcome this gap.

When $\zeta < 0$ only $\epsilon_{+;{\bf k}}^{2}$ can become less than zero and cause instability of the system. For example, close to the ${\bm \Gamma}$ point, we expand $\vert \gamma_{\bf k}\vert \approx 3- \frac{k^2}{4}$ and obtain for the threshold $k_{\mathrm{th}}=\sqrt{\frac{\Omega}{SJ}}\frac{\Delta^2}{SJ}$ of the instability. When $3SJ>\zeta > 0$ only $\epsilon_{-;{\bf k}}^{2}$ can become less than zero. Performing the same approximations as for the $\zeta < 0$ case, we get for the threshold value of the momentum $k_{\mathrm{th}} = \frac{6\Delta^2}{\sqrt{SJ(3SJ - \zeta)}}$ for $3SJ - \zeta > \frac{3}{2}\Delta^2$, i.e. away from the ${\bm \Gamma}$ point, and $k_{\mathrm{th}} = \frac{2\sqrt{6}\Delta}{\sqrt{SJ}}$ for $3SJ - \zeta < \frac{3}{2}\Delta^2$ - in the vicinity of the ${\bf \Gamma}$ point. Rescattering processes shown in Fig. \ref{fig:vertex}B result for the $3SJ>\zeta >0$ case in 
\begin{align}
\Delta^2 \rightarrow \frac{\Delta^2}{1+ \frac{3}{16\pi S} \ln\left(\frac{SJ \Lambda^2 }{4(3SJ - \zeta)} \right) + i\frac{3}{16S}},
\end{align}
where $\Lambda$ is the high-frequency cut-off.
Therefore, as $\Omega \rightarrow 6SJ$, the pairing of magnons vanishes. 
We think that this might be natural, as the one pump quanta absorption is the most effective at $\Omega = 6SJ$, and the two pump quanta absorption channel must thus get closed.

In addition to studied rescattering processes, one needs to include magnon decay rate, which originates due to interactions in second-order perturbation theory, to the main diagonal in the secular equation Eq. (\ref{secular}). Then the threshold value is going to be decreased by the decay rate.
In particular, \cite{Pershoguba} showed (see Fig. 2 there) that the decay rate, which is $\frac{1}{\tau} \propto T^{2}$, for a honeycomb lattice ferromagnet is the smallest for the Dirac magnons and is the largest for the $\varepsilon_{+;{\bf k}}$ magnons in the vicinity of the ${\bm \Gamma}$ point. 
Therefore, the threshold value for the Dirac magnon paired state instability does not get drastically modified by the decay rate.
All in all, from Eq. (\ref{ladder}) and discussions above we conclude that the Dirac magnon paired state can become the most unstable for temperatures below in vicinity of the Curie temperature.

There are two corollaries which can be made on the nature of the Dirac magnon paired state. 
First of all, when the conditions for the instability are met (divergence of Eq. (\ref{ladder})), there is going to be an absorption peak at frequency equal to $\Omega = 3SJ$ corresponding to the Dirac magnon paired state.
If the system is finite and isolated, the exponential growth of the Dirac magnons pairs in time can't last forever, and it will be stopped by interactions between the magnons, effects which are beyond studied in the present letter. 
Secondly, it can be deduced that the Dirac magnon paired state is the instability of a ferromagnet towards formation of a zero or reduced magnetization state. 
The energy to flip one spin in the unit cell equals $6S^2J$, and absorption of a pair of Dirac magnons corresponds to $6SJ$ energy. 
Then, for integer spins it is possible to flip the spin by creating appropriate number of Dirac magnon pairs. For example, for $S=1$ absorption of one pair of Dirac magnons flips the spin, for $S=2$ it is four pairs of Dirac magnons, and so on. For half-odd-integer spins, the magnetization can't be reduced to zero.
To start thinking about such state, one can imagine dynamically generated antiferromagnetic order on the honeycomb lattice. 
However, such antiferromagnetic order will be fluctuating in time between different configurations with zero magnetization. 
Unlike in the experiment \cite{Demokritov}, we restrain ourselves from calling the Dirac magnon paired state as the Bose-Einstein condensate (BEC) of Dirac magnons, instead, we hypothesise that it is a condensate of Dirac magnon pairs.
Detailed understanding of the nature of the new state is a question for future research.

In passing, let us discuss another possiblity of pumping the magnons. First note that in the honeycomb lattice there are two energy branches at the ${\bm \Gamma}$ point corresponding to $\epsilon_{+;0} = 6SJ$ and $\epsilon_{-;0} = 0$, which are connected by $\Omega = 6SJ$ frequency.
Therefore, one can excite a single magnon by a pump Eq. (\ref{pump}) with a frequency via a $\epsilon_{-;0} + \Omega \rightarrow \epsilon_{+;0}$ process.
Pumping a single magnon will not make the system unstable in a sense of Eqs. (\ref{secular}) and (\ref{pair}). However, an additional rescattering of the excited magnon with frequency $\epsilon_{+;0}$ in to a pair of Dirac magnons, via $\epsilon_{+;0} + \epsilon_{-;0} \rightarrow \epsilon_{+;{\bf K}} +  \epsilon_{-;{\bf K}^\prime}$ (schematically) processes, might create the Dirac magnons paired state and may cause an instability in the system.
This pumping scheme is the parametric pumping similar to the one in the experiment \cite{Demokritov}. It is possible that the Dirac magnon paired state is also going to occur in such pumping scheme. 
However,  it is going to coexist with unpaired magnons at $\varepsilon_{+;{\bf k}} = 6SJ$ and $\varepsilon_{-;{\bf k}} = 0$ energy. This is another question for future research.

To conclude, we studied second-order Suhl processes in a honeycomb ferromagnet and showed that under certain conditions the resonant pump's frequency corresponds to the energy of the Dirac points, causing an instability of the ferromagnet. This is because, as is schematically shown in Fig. \ref{fig:pair}B, the system can accomodate the largest ammount of magnon pairs, and their rescattering processes of the Cooper ladder type shown in Fig. \ref{fig:vertex}B result in a pole structure Eq. (\ref{ladder}), which can become singular as a function of temperature. We deduced that the instability is towards formation of zero or reduced magnetization state, and called it as the Dirac magnons paired state. We estimated that the CrBr$_{3}$ or CrCl$_{3}$ ferromagnet might show this Dirac magnon paired state below in vicinity of the Curie temperature.

{\it Acknowledgements.}
The author thanks A.M. Finkel'stein and A.Yu. Zyuzin for helpful discussions, and to Pirinem School of Theoretical Physics for hospitality. 
This work was started by the author in a research group of A.V. Balatsky in Nordita, whom the author thanks for discussions.
This work is supported by the VILLUM FONDEN via the Centre of Excellence for Dirac Materials (Grant No. 11744), the European Research Council under the European Unions Seventh Framework Program Synergy HERO, and the Knut and Alice Wallenberg Foundation KAW.

\begin{widetext}
\newpage
\setcounter{equation}{0}
\setcounter{page}{0}
\setcounter{figure}{0}

\section{Supplemental Material for "Dirac magnons pairing via pumping"}

\section{Ferromagnet on a honeycomb lattice}

\begin{figure}[h] 
\centerline{
\includegraphics[width=0.19\textwidth,height=0.14\textheight]{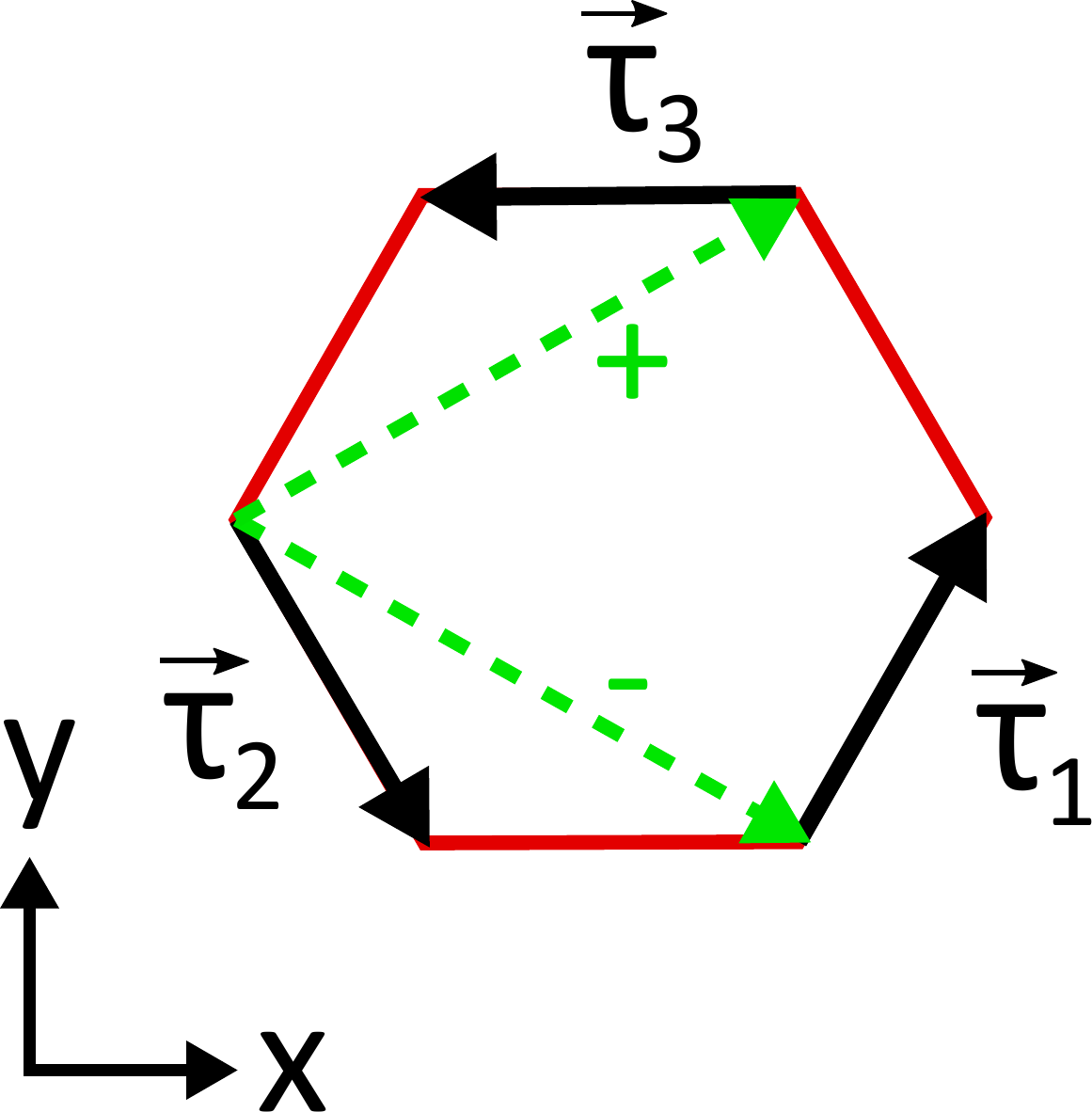}
}
\protect\caption{Schematics of the honeycomb lattice. Ferromagnetic order is assumed to be in the $z-$ direction. Vectors connecting the nearest neighbor cites are ${\bm \tau}_{1}=\frac{1}{2}\left( \frac{1}{\sqrt{3}},1\right)$, ${\bm \tau}_{2} = \frac{1}{2}\left( \frac{1}{\sqrt{3}},-1\right)$, and ${\bm \tau}_{3} = \frac{1}{\sqrt{3}}(-1,0)$. Green dashed lines correspond to the sign convention of the $\nu_{ij} = \pm 1$, which enter the Dzyaloshinskii-Moriya interaction.}

\label{fig:honey}  

\end{figure}
We study spins of the length $S$ on the honeycomb lattice. 
The spins interact via the ferromagnetic Heisenberg interaction.
We assume the order to be in $z$-direction, and wish to understand the spin waves about the order. 
We follow standard procedure discussed, for example, in books on magnetism \cite{ABP1967,Auerbach,Rezende}.
Holstein-Primakoff bosons for the spin operators
$S^{\pm} = S^{x}\pm i S^{y}$, and $S^{z}$ read
\begin{align}
S^{+} = \sqrt{2S - a^{\dag}a}a,~~~ S^{-} = a^{\dag}\sqrt{2S - a^{\dag}a},~~~ S^{z} = S - a^{\dag}a.
\end{align}
Exchange interaction is 
\begin{align}
H_{\mathrm{ex}} 
= -J\sum_{\langle ij \rangle} \left( S^{x}_{i}S^{x}_{j} + S^{y}_{i}S^{y}_{j}+ S^{z}_{i}S^{z}_{j} \right)
= -J\sum_{\langle ij \rangle} \left(\frac{1}{2} S^{+}_{i}S^{-}_{j} + \frac{1}{2} S^{-}_{i}S^{+}_{j}+ S^{z}_{i}S^{z}_{j} \right),
\end{align}
where $\langle .. \rangle$ stands for the nearest-neighbor interaction.
We are assuming $S>1$ so that $\frac{1}{S}$ expansion applies. 
This allows us to drop out higher orders of interaction between magnons.
Hamiltonian of interacting spin-waves reads,
\begin{align}
H_{\mathrm{sw}} = 
&
- JS \sum_{\langle ij \rangle} \left( a_{i}^{\dag}b_{j} + b_{j}^{\dag}a_{i} \right)
+3JS \sum_{\langle ij \rangle} \left( a_{i}^{\dag}a_{i} +  b_{j}^{\dag}b_{j} \right)
\\
&
+ \frac{J}{4}\sum_{\langle ij \rangle} a_{i}^{\dag}a_{i}a_{i}b_{j}^{\dag}
+ \frac{J}{4}\sum_{\langle ij \rangle} a_{i}b_{j}^{\dag}b_{j}^{\dag}b_{j}
+ \frac{J}{4}\sum_{\langle ij \rangle} a_{i}^{\dag}a_{i}^{\dag}a_{i}b_{j}
+ \frac{J}{4}\sum_{\langle ij \rangle} a_{i}^{\dag}b_{j}^{\dag}b_{j}b_{j}
-J\sum_{\langle ij \rangle}a_{i}^{\dag}a_{i} b_{j}^{\dag}b_{j}.
\end{align}
Fourier transform of the Hamiltonian reads as
\begin{align}
&
H_{\mathrm{sw}} \approx
- JS \int_{\bf k} 
\left( 
\gamma_{{\bf k}} a_{{\bf k}}^{\dag}b_{{\bf k}} 
+ 
\gamma_{{\bf k}}^{*} b_{{\bf k}}^{\dag}a_{{\bf k}} \right)
+3JS \int_{\bf k} 
\left( 
a_{{\bf k}}^{\dag}a_{{\bf k}} +  b_{{\bf k}}^{\dag}b_{{\bf k}} \right)
-J\int_{\{ {\bf k}\}}\delta_{{\bf k}_{1}-{\bf k}_{2},{\bf k}_{4}-{\bf k}_{3}} 
\gamma_{{\bf k}_{4}-{\bf k}_{3}} 
a_{{\bf k}_{1}}^{\dag}b_{{\bf k}_{3}}^{\dag} a_{{\bf k}_{2}} b_{{\bf k}_{4}}
\\
&
+ \frac{J}{4}\int_{\{ {\bf k}\}}
\delta_{{\bf k}_{1}-{\bf k}_{2},{\bf k}_{4}-{\bf k}_{3}} 
\left[
\gamma_{{\bf k}_{3}}^{*}
 a_{{\bf k}_{1}}^{\dag}b_{{\bf k}_{3}}^{\dag}  a_{{\bf k}_{2}}a_{{\bf k}_{4}}
+
\gamma_{{\bf k}_{3}}
 a_{{\bf k}_{2}}^{\dag}a_{{\bf k}_{4}}^{\dag} a_{{\bf k}_{1}}b_{{\bf k}_{3}}
\right]
+ \frac{J}{4}\int_{\{ {\bf k}\}}
\delta_{{\bf k}_{1}-{\bf k}_{2},{\bf k}_{4}-{\bf k}_{3}} 
\left[
\gamma_{{\bf k}_{1}}
 a_{{\bf k}_{1}}^{\dag}b_{{\bf k}_{3}}^{\dag}  b_{{\bf k}_{2}}b_{{\bf k}_{4}}
+
\gamma_{{\bf k}_{1}}^{*}
 b_{{\bf k}_{2}}^{\dag}b_{{\bf k}_{4}}^{\dag}  a_{{\bf k}_{1}}b_{{\bf k}_{3}}
\right],
\nonumber
\end{align}
where $\gamma_{\bf k} = \sum_{i=1,2,3}e^{i{\bf k}{\bm \tau}_{i}} =  2e^{i\frac{k_{x}}{2\sqrt{3}}}\cos\left( \frac{k_{y}}{2}\right)+ e^{-i\frac{k_{x}}{\sqrt{3}}}$ is the dispersion (see Fig. \ref{fig:honey} for defintions of ${\bm \tau}_{i}$ vectors) , $\{ {\bf k}\} \equiv {\bf k}_{1},{\bf k}_{2},{\bf k}_{3},{\bf k}_{4}$, and $\delta_{{\bf k}_{1},{\bf k}_{2}}\equiv 2\pi \delta({\bf k}_{1}-{\bf k}_{2})$ is the delta-function. 
Note that the two first lines of the interaction are written in the convenient for conjugation way. 
The last line is already Hermitian conjugate to itself.
The interaction is instantaneous in time.
This implies certain frequency dependence, for example,
\begin{align}
&
-J\int_{\{ {\bf k}\}}\delta_{{\bf k}_{1}-{\bf k}_{2},{\bf k}_{4}-{\bf k}_{3}} 
\int_{\epsilon_{1},\epsilon_{2},\epsilon_{3},\epsilon_{4}}
a_{\epsilon_{1};{\bf k}_{1}}^{\dag}b_{\epsilon_{3};{\bf k}_{3}}^{\dag}   
a_{\epsilon_{2};{\bf k}_{2}} b_{\epsilon_{4};{\bf k}_{4}}
\delta_{\epsilon_{1}-\epsilon_{2},\epsilon_{4}-\epsilon_{3}}
\\
=
&
-J\int_{\{ {\bf k}\}}\delta_{{\bf k}_{1}-{\bf k}_{2},{\bf k}_{4}-{\bf k}_{3}} 
\int_{\epsilon_{1},\epsilon_{3},\omega}
a_{\epsilon_{1};{\bf k}_{1}}^{\dag}
b_{\epsilon_{3};{\bf k}_{3}}^{\dag}
a_{\epsilon_{1}-\omega;{\bf k}_{2}} 
b_{\epsilon_{3}+\omega;{\bf k}_{4}}.
\end{align}


In the space of unitary cell, in which case the boson operators are defined by $\Psi_{\bf k}^\dag = ( a_{\bf k}^{\dag}, ~b_{\bf k}^{\dag})$ the Hamiltonian of linear spin-waves reads as
\begin{align}
\hat{H} = 
JS
\left[ 
\begin{array}{cc}
3 & - \gamma_{\bf k} \\
- \gamma_{\bf k}^{*} & 3
\end{array}
\right],
\end{align}
diagonalization immediatly gives energy spectrum,
\begin{align}
\epsilon_{\pm {\bf k}} =JS\left( 3 \pm  \vert \gamma_{\bf k}\vert \right)
\end{align}
with corresponding wave functions
\begin{align}
\varphi_{+} = 
\frac{1}{\sqrt{2}}
\left[ 
\begin{array}{c}
-\frac{\gamma_{\bf k}}{\vert \gamma_{\bf k} \vert} \\
1
\end{array}
\right],~~~
\varphi_{-} = 
\frac{1}{\sqrt{2}}
\left[ 
\begin{array}{c}
\frac{\gamma_{\bf k}}{\vert \gamma_{\bf k} \vert} \\
1
\end{array}
\right],
\end{align}

Green function is
\begin{align}
G_{\alpha\beta}^{\mathrm{R}/\mathrm{A}} (\epsilon,{\bf k})
= 
\frac{\varphi_{+,{\bf k}}\varphi^{\dag}_{+,{\bf k}}}{\epsilon - \epsilon_{+,{\bf k}} \pm i0}
+
\frac{\varphi_{-,{\bf k}}\varphi^{\dag}_{-,{\bf k}}}{\epsilon - \epsilon_{-,{\bf k}} \pm i0},
\end{align}
where $\alpha$ and $\beta$ are pseudospins. 
Green function can be presented in a more convenient way
\begin{align}
G_{\alpha\beta}^{\mathrm{R}/\mathrm{A}} (\epsilon,{\bf k})
=
&
\frac{1}{2}\left( \frac{1}{\epsilon - \epsilon_{+,{\bf k}} \pm i0} + \frac{1}{\epsilon - \epsilon_{-,{\bf k}} \pm i0}  \right)
-
\frac{1}{2}\left( \frac{1}{\epsilon - \epsilon_{+,{\bf k}} \pm i0} -  \frac{1}{\epsilon - \epsilon_{-,{\bf k}} \pm i0}  \right)
\left[ \begin{array}{cc}
0 & \frac{\gamma_{\bf k}}{\vert \gamma_{\bf k}\vert } \\
\frac{\gamma_{\bf k}^{*}}{\vert \gamma_{\bf k}\vert } & 0
\end{array}\right].
\end{align}


The pumping is 
\begin{align}
H_{\mathrm{pump}} 
&
= \Gamma \sum_{i} \left[ S^{x}_{i}\cos(\Omega t) + S^{y}_{i}\sin(\Omega t) \right]
= \frac{\Gamma}{2} \sum_{i} \left[ S^{+}_{i}e^{-i\Omega t} +  S^{-}_{i}e^{i\Omega t}  \right]
\\
&
\approx
\sqrt{2S}\frac{\Gamma}{2} \sum_{i} \left[ a_{i}e^{-i\Omega t} +  a^{\dag}_{i}e^{i\Omega t}  \right]
+
\sqrt{2S}\frac{\Gamma}{2} \sum_{i} \left[ b_{i}e^{-i\Omega t} +  b^{\dag}_{i}e^{i\Omega t}  \right].
\end{align}

For the sake of discussion, we also consider Dzyaloshinskii-Moriya interaction
\begin{align}\label{DMI_Hamiltonian}
H_{\mathrm{DMI}} = D \sum_{\langle\langle ij \rangle\rangle}\nu_{ij}[{\bf S}_{i}\times{\bf S}_{j}]_{z},
\end{align}
where $\langle\langle ij \rangle\rangle$ stands for the next-nearest neighbor interaction, and $\nu_{ij}=\pm 1$ depending on the direction of interaction with the signs defined by green dashed arrows in Fig. (\ref{fig:honey}).   
In Holstein-Primakoff boson representation of spins, the DMI becomes 
\begin{align}
H_{\mathrm{DMI}} = SD\int_{\bf k} \xi_{\bf k} \left(a^{\dag}_{\bf k}a_{\bf k} -  b^{\dag}_{\bf k}b_{\bf k} \right),
\end{align}
where $\xi_{\bf k} =2\left[ \sin(k_{y}) - 2\sin\left( \frac{k_{y}}{2}\right)\cos\left(\frac{\sqrt{3}k_{x}}{2} \right)\right]$.

\section{Keldysh formalism}
We stress that in the hindsight, the Keldysh technique is certainly not the only choice for the problem at hand. It seems that Matsubara frequency space should work equally well. However, as the system under study is pumped and formally out-of-equilibrium, we decided to be on a safe side and follow non-equilibrium field theory technique - the Keldysh technique.
Here we briefly outline steps of the Keldysh technique, which we utilized in analysis of the system. 
For a detailed review of the Keldysh formalism see book \cite{Kamenev}, which is going to be followed below.
When considering the action of non-interacting magnons, the integral over the Keldysh contour is split as usual in to forward $\bar{\Psi}^{+}, \Psi^{+}$ and backward $\bar{\Psi}^{-},\Psi^{-}$ parts. For example, a part containing non-interacting Hamiltonian transforms as
\begin{align}
\int_{{\cal C}}dt \bar{\Psi}(t)\hat{H}\Psi (t)= 
\int_{-\infty}^{+\infty}dt
\bar{\Psi}^{+}(t)\hat{H} \Psi^{+} (t)
- 
\int_{-\infty}^{+\infty}dt
\bar{\Psi}^{-}(t)\hat{H} \Psi^{-}(t)
=  
\int_{-\infty}^{+\infty}dt 
\left[
\bar{\Psi}^{\mathrm{cl}}(t)\hat{H}\Psi^{\mathrm{q}}(t)
+
\bar{\Psi}^{\mathrm{q}}(t)\hat{H}\Psi^{\mathrm{cl}}(t)
\right],
\end{align}
where 
\begin{align}
\Psi^{\mathrm{cl}/\mathrm{q}} = \frac{1}{\sqrt{2}}\left( \Psi^{+} \pm \Psi^{-} \right),
\end{align}
and the same for $\bar{\Psi}$ fields.
The action of non-interacting magnons is
\begin{align}
iS = 
i\int_{-\infty}^{+\infty}dt 
\bar{\Psi}(t)
\left[
\begin{array} {cc}
0 &   \left[G^{-1} \right]^{\mathrm{A}} \\
\left[ G^{-1} \right]^{\mathrm{R}} & \left[ G^{-1} \right]^{\mathrm{K}}
\end{array}
\right]
\Psi(t)
\end{align}
where 
\begin{align}
\Psi = \left[ \begin{array}{c} \Psi^{\mathrm{cl}} \\ \Psi^{\mathrm{q}}\end{array}\right], ~~~~
\bar{\Psi} = \left[ \begin{array}{cc} \bar{\Psi}^{\mathrm{cl}} & \bar{\Psi}^{\mathrm{q}}\end{array}\right],
\end{align}
and $ \left[G^{-1}(\epsilon) \right]^{\mathrm{R}/\mathrm{A}} = \epsilon \pm i0 - \hat{H}$ is the inverse Green function in the Fourier space.
Note that the $[G^{-1}]^{\mathrm{K}}$ is the quantum-quantum component of the action, and the classical-classical component of the action is absent.
The Green function is 
\begin{align}
\langle \Psi(t) \bar{\Psi}(t^\prime) \rangle_{S} = i
\left[
\begin{array} {cc}
G^{\mathrm{K}}(t-t^\prime) &  G^{\mathrm{R}}(t-t^\prime)  \\
G^{\mathrm{A}}(t-t^\prime) &  0
\end{array}
\right],
\end{align}
where in particular 
\begin{align}
&
\langle \Psi^{\mathrm{cl}}(t) \bar{\Psi}^{\mathrm{cl}}(t^\prime) \rangle_{S} =\sum_{\epsilon} i G^{\mathrm{K}}(\epsilon) e^{-i\epsilon (t-t^\prime)},
\\
&
\langle \Psi^{\mathrm{cl}}(t) \bar{\Psi}^{\mathrm{q}}(t^\prime) \rangle_{S} =\sum_{\epsilon} i G^{\mathrm{R}}(\epsilon) e^{-i\epsilon (t-t^\prime)},
\\
&
\langle \Psi^{\mathrm{q}}(t) \bar{\Psi}^{\mathrm{cl}}(t^\prime) \rangle_{S} = \sum_{\epsilon} i G^{\mathrm{A}}(\epsilon) e^{-i\epsilon (t-t^\prime)}.
\end{align}
In frequency space
\begin{align}
&
\langle \Psi^{\mathrm{cl}}(\epsilon_{1}) \bar{\Psi}^{\mathrm{cl}}(\epsilon_{2}) \rangle_{S} 
= i G^{\mathrm{K}}(\epsilon_{1}) \delta_{\epsilon_{1},\epsilon_{2}},
\\
&
\langle \Psi^{\mathrm{cl}}(\epsilon_{1}) \bar{\Psi}^{\mathrm{q}}(\epsilon_{2}) \rangle_{S} =
 i G^{\mathrm{R}}(\epsilon_{1}) \delta_{\epsilon_{1},\epsilon_{2}},
\\
&
\langle \Psi^{\mathrm{q}}(\epsilon_{1}) \bar{\Psi}^{\mathrm{cl}}(\epsilon_{2}) \rangle_{S} 
=  i G^{\mathrm{A}}(\epsilon_{1})  \delta_{\epsilon_{1},\epsilon_{2}},
\end{align}
where $\delta_{\epsilon_{1},\epsilon_{2}} = 2\pi\delta(\epsilon_{1}-\epsilon_{2})$ is the delta-function.
The Green function must satisfy unity identity (here everywhere multiplication assumes convolution in time),
\begin{align}
\left[
\begin{array} {cc}
0 &   \left[G^{-1} \right]^{\mathrm{A}} \\
\left[ G^{-1} \right]^{\mathrm{R}} & \left[ G^{-1} \right]^{\mathrm{K}}
\end{array}
\right]
\left[
\begin{array} {cc}
G^{\mathrm{K}} &  G^{\mathrm{R}}  \\
G^{\mathrm{A}} &  0
\end{array}
\right] = 1,
\end{align}
which gives us a condition on $G^{\mathrm{K}}$ function
\begin{align}
 \left[G^{-1} \right]^{\mathrm{R}} G^{\mathrm{K}} + \left[ G^{-1} \right]^{\mathrm{K}} G^{\mathrm{A}} = 0,
\end{align}
which means
\begin{align}
\left[ G^{-1} \right]^{\mathrm{K}} = - \left[G^{-1} \right]^{\mathrm{R}} G^{\mathrm{K}} \left[G^{-1} \right]^{\mathrm{A}}.
\end{align}
With the parametrization 
\begin{align}
G^{\mathrm{K}}  = G^{\mathrm{R}}  {\cal F} - {\cal F}  G^{\mathrm{A}}, 
\end{align}
where ${\cal F}$ is the distribution function, we get
\begin{align}
\left[ G^{-1} \right]^{\mathrm{K}} 
= \left[G^{-1} \right]^{\mathrm{R}} {\cal F} -  {\cal F} \left[G^{-1} \right]^{\mathrm{A}}.
\end{align}
This is the kinetic equation determining distribution function. 

The pumping field is described by 
\begin{align}
\frac{\sqrt{2S}\Gamma}{2}
\int_{\cal C}dt \left( \Psi e^{-i\Omega t} + \bar{\Psi} e^{i\Omega t} \right) 
= 
\Gamma \sqrt{S} \int_{-\infty}^{+\infty}dt \left( \Psi^{\mathrm{q}} e^{-i\Omega t} + \bar{\Psi}^{\mathrm{q}} e^{i\Omega t}  \right).
\end{align}
This might update the Hamiltonian and the Green functions. 
To check this, we can use the following identity, 
\begin{align}
\int d[\bar{\Psi},\Psi] e^{ - \sum_{ij} \bar{\Psi}_{i}\hat{A}_{ij}\Psi_{j}  + \sum_{i}\left( \bar{\Psi}_{i}J_{i} +  \bar{J}_{i}\Psi_{i} \right)   } = \frac{1}{\mathrm{det}\hat{A}}e^{\sum_{ij} \bar{J}_{i}(\hat{A}^{-1})_{ij}J_{j}}
\end{align}
and since there is no $\mathrm{q}$-$\mathrm{q}$ element in the $\hat{A}^{-1}$ matrix, the pumping field will not enter the final result of integration. 
However, the corresponding classical fields and consequently Green functions are going to be affected by the pumping fields. We are going to go over that in the next subsection.

Now let us include interactions between magnons.
Schematically, general four-boson interaction rewritten in terms of Keldysh fields is 
\begin{align}
\int_{{\cal C}} dt \bar{\Psi}_{1}\Psi_{2}\bar{\Psi}_{3}\Psi_{4}
&
 =
\int_{-\infty}^{+\infty}dt
 \bar{\Psi}_{1}^{+}\Psi_{2}^{+}\bar{\Psi}_{3}^{+}\Psi_{4}^{+}
- 
\int_{-\infty}^{+\infty}dt
 \bar{\Psi}_{1}^{-}\Psi_{2}^{-}\bar{\Psi}_{3}^{-}\Psi_{4}^{-}
\\
&
=
\frac{1}{2}
\int_{-\infty}^{+\infty}dt
\left(\bar{\Psi}_{1}^{\mathrm{cl}}\Psi_{2}^{\mathrm{cl}}  + \bar{\Psi}_{1}^{\mathrm{q}}\Psi_{2}^{\mathrm{q}} \right)
\left(\bar{\Psi}_{3}^{\mathrm{cl}}\Psi_{4}^{\mathrm{q}}  + \bar{\Psi}_{3}^{\mathrm{q}}\Psi_{4}^{\mathrm{cl}}\right) 
\\
&
+
\frac{1}{2}
\int_{-\infty}^{+\infty}dt
\left(\bar{\Psi}_{1}^{\mathrm{cl}}\Psi_{2}^{\mathrm{q}}  + \bar{\Psi}_{1}^{\mathrm{q}}\Psi_{2}^{\mathrm{cl}} \right)
\left(\bar{\Psi}_{3}^{\mathrm{cl}}\Psi_{4}^{\mathrm{cl}}  + \bar{\Psi}_{3}^{\mathrm{q}}\Psi_{4}^{\mathrm{q}}\right)
\end{align}
where $1,2,3,4$ indeces stand for a general frequency-momentum-spin variable. 
Under relabelling, the two terms after second equality sign double each other, but for the sake of generality kept as they are.

\subsection{Shifting the pump field away}
Lagrangian describing non-interacting magnons with the pump's frequency $\Omega$ and momentum ${\bf k} = 0$ is schematically written as
\begin{align}
{\cal L}_{0,\Omega}
&
=
\sum_{m,n}\bar{\Psi}^{\mathrm{q}}_{m, 0,\Omega} \hat{{\cal L}}^{\mathrm{K}}_{mn,0,\Omega} \Psi^{\mathrm{q}}_{n, 0,\Omega}
+\sum_{m,n}\bar{\Psi}^{\mathrm{cl}}_{m, 0, \Omega} \hat{{\cal L}}^{\mathrm{A}}_{mn,0,\Omega} \Psi^{\mathrm{q}}_{n, 0, \Omega} + \sum_{m,n}\bar{\Psi}^{\mathrm{q}}_{m, 0, \Omega} \hat{{\cal L}}^{\mathrm{R}}_{mn,0,\Omega} \Psi^{\mathrm{cl}}_{n, 0, \Omega}
\\
&
-\Gamma\sqrt{S}\sum_{n} \Psi^{\mathrm{q}}_{n, 0, \Omega} 
-\Gamma\sqrt{S}\sum_{n} \bar{\Psi}^{\mathrm{q}}_{n, 0, \Omega},
\end{align}
where $\hat{{\cal L}}^{\mathrm{K}/\mathrm{R}/\mathrm{A}}_{mn,0,\Omega}$ is the Lagrangian density corresponding to Keldysh, retarded or advanced part correspondingly. For example $\hat{{\cal L}}^{\mathrm{R}/\mathrm{A}}_{mn,{\bf k},\Omega} = (\Omega \pm i0)\delta_{mn} - [\hat{H}_{0}]_{mn,{\bf k}}$.
The advanced part of the Lagrangian is 
\begin{align}
{\cal L}_{0,\Omega}^{\mathrm{A}}=
\sum_{m,n}\bar{\Psi}^{\mathrm{cl}}_{m, 0, \Omega} \hat{{\cal L}}^{\mathrm{A}}_{mn,0,\Omega} \Psi^{\mathrm{q}}_{n, 0, \Omega}
-\Gamma\sqrt{S}\sum_{n} \Psi^{\mathrm{q}}_{n, 0, \Omega}  ,
\end{align}
in which we would like to shift away terms linear in $\Psi^{\mathrm{q}}_{n, 0, \Omega}$. 
We achieve it with
\begin{align}
&
\bar{\Psi}^{\mathrm{cl}}_{\alpha, 0, \Omega} \rightarrow \bar{\Psi}^{\mathrm{cl}}_{\alpha, 0, \Omega} + x_{\mathrm{A}}, 
\\
&
\bar{\Psi}^{\mathrm{cl}}_{\beta, 0, \Omega} \rightarrow \bar{\Psi}^{\mathrm{cl}}_{\beta, 0, \Omega} + y_{\mathrm{A}}, 
\end{align}
with  
\begin{align}
&
x_{\mathrm{A}}= \frac{{\cal L}^{\mathrm{A}}_{\beta\alpha,0,\Omega} - {\cal L}^{\mathrm{A}}_{\beta\beta,0,\Omega} }{ {\cal L}^{\mathrm{A}}_{\alpha\beta,0,\Omega}{\cal L}^{\mathrm{A}}_{\beta\alpha,0,\Omega}   - {\cal L}^{\mathrm{A}}_{\beta\beta,0,\Omega}{\cal L}^{\mathrm{A}}_{\alpha\alpha,0,\Omega}}\Gamma\sqrt{S},
\\
&
y_{\mathrm{A}}= \frac{{\cal L}^{\mathrm{A}}_{\alpha\beta,0,\Omega} - {\cal L}^{\mathrm{A}}_{\alpha\alpha,0,\Omega} }{ {\cal L}^{\mathrm{A}}_{\alpha\beta,0,\Omega}{\cal L}^{\mathrm{A}}_{\beta\alpha,0,\Omega}   - {\cal L}^{\mathrm{A}}_{\beta\beta,0,\Omega}{\cal L}^{\mathrm{A}}_{\alpha\alpha,0,\Omega}}\Gamma\sqrt{S}.
\end{align}
For the retarded analog of the Lagrangian,
\begin{align}
{\cal L}_{0,\Omega}^{\mathrm{R}}=
\sum_{m,n}\bar{\Psi}^{\mathrm{q}}_{m, 0, \Omega} {\cal L}^{\mathrm{R}}_{mn,0,\Omega} \Psi^{\mathrm{cl}}_{n, 0, \Omega}
-\Gamma\sqrt{S}\sum_{n} \bar{\Psi}^{\mathrm{q}}_{n, 0, \Omega}  ,
\end{align}
in which we would like to shift away terms linear in $\bar{\Psi}^{\mathrm{q}}_{n, 0, \Omega}$. We achieve it with
\begin{align}
&
\Psi^{\mathrm{cl}}_{\alpha, 0, \Omega} \rightarrow \Psi^{\mathrm{cl}}_{\alpha, 0, \Omega} + x_{\mathrm{R}}, 
\\
&
\Psi^{\mathrm{cl}}_{\beta, 0, \Omega} \rightarrow \Psi^{\mathrm{cl}}_{\beta, 0, \Omega} + y_{\mathrm{R}}, 
\end{align}
with  
\begin{align}
&
x_{\mathrm{R}}= \frac{{\cal L}^{\mathrm{R}}_{\beta\alpha,0,\Omega} - {\cal L}^{\mathrm{R}}_{\beta\beta,0,\Omega} }{ {\cal L}^{\mathrm{R}}_{\alpha\beta,0,\Omega}{\cal L}^{\mathrm{R}}_{\beta\alpha,0,\Omega}   - {\cal L}^{\mathrm{R}}_{\beta\beta,0,\Omega}{\cal L}^{\mathrm{R}}_{\alpha\alpha,0,\Omega}}\Gamma\sqrt{S},
\\
&
y_{\mathrm{R}}= \frac{{\cal L}^{\mathrm{R}}_{\alpha\beta,0,\Omega} - {\cal L}^{\mathrm{R}}_{\alpha\alpha,0,\Omega} }{ {\cal L}^{\mathrm{R}}_{\alpha\beta,0,\Omega}{\cal L}^{\mathrm{R}}_{\beta\alpha,0,\Omega}   - {\cal L}^{\mathrm{R}}_{\beta\beta,0,\Omega}{\cal L}^{\mathrm{R}}_{\alpha\alpha,0,\Omega}}\Gamma\sqrt{S}.
\end{align}

\section{Pumping to the Dirac points with a $\Omega = 3SJ$ frequency pump}

\subsection{Two quanta pumping to Dirac points}
Here we discuss off-resonance pumping, when the frequency of the pump is half the band-width, namely $\Omega = 3SJ$. 
There are no mass-shell states with ${\bf k} = 0$ at this frequency. 
Thus, there is no possibility to pump single magnon to this point, but due to the interactions, there is a possibility to pump a pair of magnons. See Fig. \ref{fig:pump} for the schematics of the process of absorption of two pump field quanta.
This processes is known in the literature as the second-order Suhl process \cite{Suhl1957,Rezende}
One can see it by absorbing the pumping field by shifting corresponding classical (only) fields,
\begin{align}
&
\bar{\Psi}^{\mathrm{cl}}_{\alpha, 0, \Omega} \rightarrow \bar{\Psi}^{\mathrm{cl}}_{\alpha, 0, \Omega} - \frac{\Gamma\sqrt{S}}{3SJ},
\\
&
\Psi^{\mathrm{cl}}_{\alpha, 0, \Omega} \rightarrow \Psi^{\mathrm{cl}}_{\alpha, 0, \Omega} - \frac{\Gamma\sqrt{S}}{3SJ}.
\end{align}
The shift means that a physical state with corresponding quantum numbers acquires a classical value. 
For example, if it was a Bose-Einstein condensate we were talking about, it would mean that the magnon accumulate in the state.
However, since the shifted state is off-shell, one would not expect any magnon accumulation in it.
Instead, the magnons can rescatter from this virtual state to the on-shell states according to the frequency and momentum conservation. 
To describe these effects, we notice that the interaction part of the action will be affected by the shift. 
\begin{align}
&
 iS_{\mathrm{interaction};1}=
-i\frac{J}{4}
\int_{\{\omega\}}
\int_{\{{\bf k}\}}
\gamma_{{\bf k}_{4}}
 a_{{\bf k}_{1}}^{\dag}a_{{\bf k}_{2}}a_{{\bf k}_{3}}^{\dag}b_{{\bf k}_{4}}
\delta_{{\bf k}_{1}-{\bf k}_{2},{\bf k}_{4}-{\bf k}_{3}} 
\delta_{\omega_{1}-\omega_{2},\omega_{4}-\omega_{3}}
\\
&
\rightarrow
-i\frac{J}{8}
\int_{\{\omega\}}
\int_{\{{\bf k}\}}
\gamma_{{\bf k}_{4}}
\bigg(
\bar{\Psi}^{\mathrm{cl}}_{\alpha;{\bf k}_{1};\omega_{1}} 
\bar{\Psi}^{\mathrm{cl}}_{\alpha;{\bf k}_{3};\omega_{3}}
\Psi^{\mathrm{cl}}_{\alpha;{\bf k}_{2};\omega_{2}}
\Psi^{\mathrm{q}}_{\beta;{\bf k}_{4};\omega_{4}} 
+
\bar{\Psi}^{\mathrm{cl}}_{\alpha;{\bf k}_{1};\omega_{1}} 
\bar{\Psi}^{\mathrm{q}}_{\alpha;{\bf k}_{3};\omega_{3}}
\Psi^{\mathrm{cl}}_{\alpha;{\bf k}_{2};\omega_{2}}
\Psi^{\mathrm{cl}}_{\beta;{\bf k}_{4};\omega_{4}} 
\nonumber
\\
&
+
\bar{\Psi}^{\mathrm{cl}}_{\alpha;{\bf k}_{1};\omega_{1}} 
\bar{\Psi}^{\mathrm{cl}}_{\alpha;{\bf k}_{3};\omega_{3}}
\Psi^{\mathrm{q}}_{\alpha;{\bf k}_{2};\omega_{2}}
\Psi^{\mathrm{cl}}_{\beta;{\bf k}_{4};\omega_{4}} 
+
\bar{\Psi}^{\mathrm{q}}_{\alpha;{\bf k}_{1};\omega_{1}} 
\bar{\Psi}^{\mathrm{cl}}_{\alpha;{\bf k}_{3};\omega_{3}}
\Psi^{\mathrm{cl}}_{\alpha;{\bf k}_{2};\omega_{2}}
\Psi^{\mathrm{cl}}_{\beta;{\bf k}_{4};\omega_{4}} 
\bigg)
\delta_{{\bf k}_{1}-{\bf k}_{2},{\bf k}_{4}-{\bf k}_{3}} 
\delta_{\omega_{1}-\omega_{2},\omega_{4}-\omega_{3}}
\nonumber
\\
&
-i
\frac{J}{8}
\int_{\{\omega\}}
\int_{\{{\bf k}\}}
\gamma_{{\bf k}_{4}}
\left(\frac{\Gamma \sqrt{S}}{3SJ}\right)^2
\Psi^{\mathrm{cl}}_{\alpha;{\bf k}_{2};\omega_{2}}
\Psi^{\mathrm{q}}_{\beta;{\bf k}_{4};\omega_{4}} 
\delta_{-{\bf k}_{2},{\bf k}_{4}} 
\delta_{\Omega-\omega_{2},\omega_{4}-\Omega}
\\
&
-i
\frac{J}{8}
\int_{\{\omega\}}
\int_{\{{\bf k}\}}
\gamma_{0}
\left(\frac{\Gamma \sqrt{S}}{3SJ}\right)^2
\bar{\Psi}^{\mathrm{cl}}_{\alpha;{\bf k}_{1};\omega_{1}} 
\bar{\Psi}^{\mathrm{q}}_{\alpha;{\bf k}_{3};\omega_{3}}
\delta_{{\bf k}_{1},-{\bf k}_{3}}
\delta_{\omega_{1}-\Omega,\Omega-\omega_{3}}
\nonumber
\\
&
-i
\frac{J}{8}
\int_{\{\omega\}}
\int_{\{{\bf k}\}}
\gamma_{{\bf k}_{4}}
\left(\frac{\Gamma \sqrt{S}}{3SJ}\right)^2
\Psi^{\mathrm{q}}_{\alpha;{\bf k}_{2};\omega_{2}}
\Psi^{\mathrm{cl}}_{\beta;{\bf k}_{4};\omega_{4}} 
\delta_{-{\bf k}_{2},{\bf k}_{4}} 
\delta_{\Omega-\omega_{2},\omega_{4}-\Omega}
\\
&
-i
\frac{J}{8}
\int_{\{\omega\}}
\int_{\{{\bf k}\}}
\gamma_{0}
\left(\frac{\Gamma \sqrt{S}}{3SJ}\right)^2
\bar{\Psi}^{\mathrm{q}}_{\alpha;{\bf k}_{1};\omega_{1}} 
\bar{\Psi}^{\mathrm{cl}}_{\alpha;{\bf k}_{3};\omega_{3}}
\delta_{{\bf k}_{1},-{\bf k}_{3}}
\delta_{\omega_{1}-\Omega,\Omega-\omega_{3}}.
\nonumber
\end{align}
Regarding cubic terms, in experimentally relevant limit of $\left(\frac{\Gamma \sqrt{S}}{3SJ}\right)^2 < 1$ they can be ignored.
They will contribute to the interaction between magnons, but will have $\left(\frac{\Gamma \sqrt{S}}{3SJ}\right)^2 < 1$ small factor as compared to the original interaction.
It is not possible to generate $\propto \bar{\Psi}^{\mathrm{q}}_{\alpha;{\bf k}_{1};\omega_{1}} 
\Psi^{\mathrm{cl}}_{\alpha;{\bf k}_{2};\omega_{2}}$ or $\propto \bar{\Psi}^{\mathrm{q}}_{\alpha;{\bf k}_{1};\omega_{1}} 
\Psi^{\mathrm{cl}}_{\beta;{\bf k}_{2};\omega_{2}}$ or other similar terms as they all sum up to zero. 
This cancellation occurs between all terms in the interaction (between $\propto -J$ and $\propto \frac{J}{4}$ terms in the interaction). We give an example of such cancellation in the end of this subsection.

\begin{figure}[h] 
\centerline{
\includegraphics[width=0.18\textwidth,height=0.08\textheight]{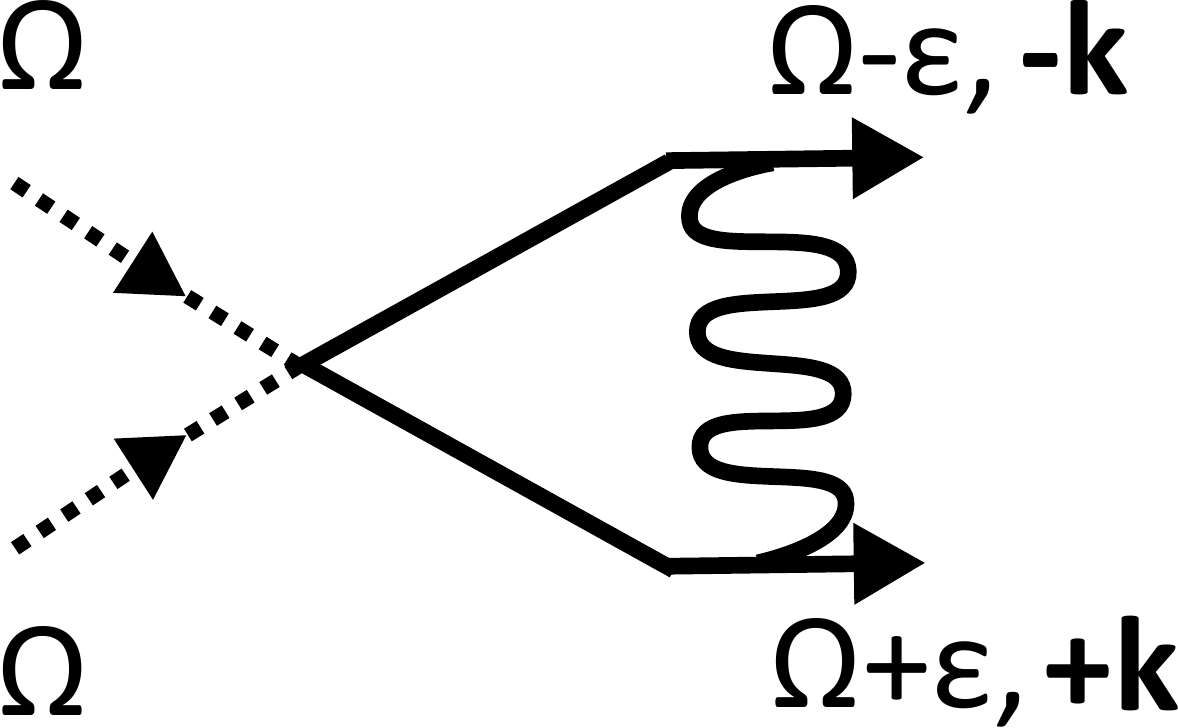}
}
\protect\caption{Schematics of two pump quanta absoroption. Here the dashed lines correspond to the pump field, while the wavy line to interaction between the magnons.}

\label{fig:pump}  

\end{figure}

Below we list four remaining terms in the interaction. 
\begin{align}
&
 iS_{\mathrm{interaction};2}=
-i\frac{J}{4}
\int_{\{\omega\}}
\int_{\{{\bf k}\}}
\gamma_{{\bf k}_{3}}^{*}
 a_{{\bf k}_{1}}^{\dag}a_{{\bf k}_{2}}b_{{\bf k}_{3}}^{\dag}a_{{\bf k}_{4}}
\delta_{{\bf k}_{1}-{\bf k}_{2},{\bf k}_{4}-{\bf k}_{3}} 
\delta_{\omega_{1}-\omega_{2},\omega_{4}-\omega_{3}}
\\
&
\rightarrow
-i\frac{J}{8}
\int_{\{\omega\}}
\int_{\{{\bf k}\}}
\gamma_{{\bf k}_{3}}^{*} 
\bigg(
\bar{\Psi}^{\mathrm{cl}}_{\alpha;{\bf k}_{1};\omega_{1}} 
\bar{\Psi}^{\mathrm{cl}}_{\beta;{\bf k}_{3};\omega_{3}}
\Psi^{\mathrm{cl}}_{\alpha;{\bf k}_{2};\omega_{2}}
\Psi^{\mathrm{q}}_{\alpha;{\bf k}_{4};\omega_{4}} 
+
\bar{\Psi}^{\mathrm{cl}}_{\alpha;{\bf k}_{1};\omega_{1}} 
\bar{\Psi}^{\mathrm{q}}_{\beta;{\bf k}_{3};\omega_{3}}
\Psi^{\mathrm{cl}}_{\alpha;{\bf k}_{2};\omega_{2}}
\Psi^{\mathrm{cl}}_{\alpha;{\bf k}_{4};\omega_{4}} 
\nonumber
\\
&
+
\bar{\Psi}^{\mathrm{cl}}_{\alpha;{\bf k}_{1};\omega_{1}} 
\bar{\Psi}^{\mathrm{cl}}_{\beta;{\bf k}_{3};\omega_{3}}
\Psi^{\mathrm{q}}_{\alpha;{\bf k}_{2};\omega_{2}}
\Psi^{\mathrm{cl}}_{\alpha;{\bf k}_{4};\omega_{4}} 
+
\bar{\Psi}^{\mathrm{q}}_{\alpha;{\bf k}_{1};\omega_{1}} 
\bar{\Psi}^{\mathrm{cl}}_{\beta;{\bf k}_{3};\omega_{3}}
\Psi^{\mathrm{cl}}_{\alpha;{\bf k}_{2};\omega_{2}}
\Psi^{\mathrm{cl}}_{\alpha;{\bf k}_{4};\omega_{4}} 
\bigg)
\delta_{{\bf k}_{1}-{\bf k}_{2},{\bf k}_{4}-{\bf k}_{3}} 
\delta_{\omega_{1}-\omega_{2},\omega_{4}-\omega_{3}}
\nonumber
\\
&
-i
\frac{J}{8}
\int_{\{\omega\}}
\int_{\{{\bf k}\}}
\gamma_{0}
\left(\frac{\Gamma \sqrt{S}}{3SJ}\right)^2
\Psi^{\mathrm{cl}}_{\alpha;{\bf k}_{2};\omega_{2}}
\Psi^{\mathrm{q}}_{\alpha;{\bf k}_{4};\omega_{4}} 
\delta_{-{\bf k}_{2},{\bf k}_{4}} 
\delta_{\Omega-\omega_{2},\omega_{4}-\Omega}
\\
&
-i
\frac{J}{8}
\int_{\{\omega\}}
\int_{\{{\bf k}\}}
\gamma_{{\bf k}_{3}}^{*} 
\left(\frac{\Gamma \sqrt{S}}{3SJ}\right)^2
\bar{\Psi}^{\mathrm{cl}}_{\alpha;{\bf k}_{1};\omega_{1}} 
\bar{\Psi}^{\mathrm{q}}_{\beta;{\bf k}_{3};\omega_{3}}
\delta_{{\bf k}_{1},-{\bf k}_{3}} 
\delta_{\omega_{1}-\Omega,\Omega-\omega_{3}}
\nonumber
\\
&
-i
\frac{J}{8}
\int_{\{\omega\}}
\int_{\{{\bf k}\}}
\gamma_{0}
\left(\frac{\Gamma \sqrt{S}}{3SJ}\right)^2
\Psi^{\mathrm{q}}_{\alpha;{\bf k}_{2};\omega_{2}}
\Psi^{\mathrm{cl}}_{\alpha;{\bf k}_{4};\omega_{4}} 
\delta_{-{\bf k}_{2},{\bf k}_{4}} 
\delta_{\Omega-\omega_{2},\omega_{4}-\Omega}
\\
&
-i
\frac{J}{8}
\int_{\{\omega\}}
\int_{\{{\bf k}\}}
\gamma_{{\bf k}_{3}}^{*} 
\left(\frac{\Gamma \sqrt{S}}{3SJ}\right)^2
\bar{\Psi}^{\mathrm{q}}_{\alpha;{\bf k}_{1};\omega_{1}} 
\bar{\Psi}^{\mathrm{cl}}_{\beta;{\bf k}_{3};\omega_{3}}
\delta_{{\bf k}_{1},-{\bf k}_{3}} 
\delta_{\omega_{1}-\Omega,\Omega-\omega_{3}},
\nonumber
\end{align}
and
\begin{align}
&
 iS_{\mathrm{interaction};3}=
-i\frac{J}{4}
\int_{\{\omega\}}
\int_{\{{\bf k}\}}
\gamma_{{\bf k}_{1}}
 a_{{\bf k}_{1}}^{\dag}b_{{\bf k}_{2}}b_{{\bf k}_{3}}^{\dag}b_{{\bf k}_{4}}
\delta_{{\bf k}_{1}-{\bf k}_{2},{\bf k}_{4}-{\bf k}_{3}} 
\delta_{\omega_{1}-\omega_{2},\omega_{4}-\omega_{3}}
\\
&
\rightarrow
-i
\frac{J}{8}
\int_{\{\omega\}}
\int_{\{{\bf k}\}}
\gamma_{{\bf k}_{1}}
\bigg(
\bar{\Psi}^{\mathrm{cl}}_{\alpha;{\bf k}_{1};\omega_{1}} 
\bar{\Psi}^{\mathrm{cl}}_{\beta;{\bf k}_{3};\omega_{3}}
\Psi^{\mathrm{cl}}_{\beta;{\bf k}_{2};\omega_{2}}
\Psi^{\mathrm{q}}_{\beta;{\bf k}_{4};\omega_{4}} 
+
\bar{\Psi}^{\mathrm{cl}}_{\alpha;{\bf k}_{1};\omega_{1}} 
\bar{\Psi}^{\mathrm{q}}_{\beta;{\bf k}_{3};\omega_{3}}
\Psi^{\mathrm{cl}}_{\beta;{\bf k}_{2};\omega_{2}}
\Psi^{\mathrm{cl}}_{\beta;{\bf k}_{4};\omega_{4}} 
\nonumber
\\
&
+
\bar{\Psi}^{\mathrm{cl}}_{\alpha;{\bf k}_{1};\omega_{1}} 
\bar{\Psi}^{\mathrm{cl}}_{\beta;{\bf k}_{3};\omega_{3}}
\Psi^{\mathrm{q}}_{\beta;{\bf k}_{2};\omega_{2}}
\Psi^{\mathrm{cl}}_{\beta;{\bf k}_{4};\omega_{4}} 
+
\bar{\Psi}^{\mathrm{q}}_{\alpha;{\bf k}_{1};\omega_{1}} 
\bar{\Psi}^{\mathrm{cl}}_{\beta;{\bf k}_{3};\omega_{3}}
\Psi^{\mathrm{cl}}_{\beta;{\bf k}_{2};\omega_{2}}
\Psi^{\mathrm{cl}}_{\beta;{\bf k}_{4};\omega_{4}} 
\bigg)
\delta_{{\bf k}_{1}-{\bf k}_{2},{\bf k}_{4}-{\bf k}_{3}} 
\delta_{\omega_{1}-\omega_{2},\omega_{4}-\omega_{3}}
\nonumber
\\
&
-i
\frac{J}{8}
\int_{\{\omega\}}
\int_{\{{\bf k}\}}
\gamma_{0}
\left(\frac{\Gamma \sqrt{S}}{3SJ}\right)^2
\Psi^{\mathrm{cl}}_{\beta;{\bf k}_{2};\omega_{2}}
\Psi^{\mathrm{q}}_{\beta;{\bf k}_{4};\omega_{4}} 
\delta_{-{\bf k}_{2},{\bf k}_{4}} 
\delta_{\Omega-\omega_{2},\omega_{4}-\Omega}
\\
&
-i
\frac{J}{8}
\int_{\{\omega\}}
\int_{\{{\bf k}\}}
\gamma_{{\bf k}_{1}}
\left(\frac{\Gamma \sqrt{S}}{3SJ}\right)^2
\bar{\Psi}^{\mathrm{cl}}_{\alpha;{\bf k}_{1};\omega_{1}} 
\bar{\Psi}^{\mathrm{q}}_{\beta;{\bf k}_{3};\omega_{3}}
\delta_{{\bf k}_{1},-{\bf k}_{3}}
\delta_{\omega_{1}-\Omega,\Omega-\omega_{3}}
\nonumber
\\
&
-i
\frac{J}{8}
\int_{\{\omega\}}
\int_{\{{\bf k}\}}
\gamma_{0}
\left(\frac{\Gamma \sqrt{S}}{3SJ}\right)^2
\Psi^{\mathrm{q}}_{\beta;{\bf k}_{2};\omega_{2}}
\Psi^{\mathrm{cl}}_{\beta;{\bf k}_{4};\omega_{4}} 
\delta_{-{\bf k}_{2},{\bf k}_{4}} 
\delta_{\Omega-\omega_{2},\omega_{4}-\Omega}
\\
&
-i
\frac{J}{8}
\int_{\{\omega\}}
\int_{\{{\bf k}\}}
\gamma_{{\bf k}_{1}}
\left(\frac{\Gamma \sqrt{S}}{3SJ}\right)^2
\bar{\Psi}^{\mathrm{q}}_{\alpha;{\bf k}_{1};\omega_{1}} 
\bar{\Psi}^{\mathrm{cl}}_{\beta;{\bf k}_{3};\omega_{3}}
\delta_{{\bf k}_{1},-{\bf k}_{3}}
\delta_{\omega_{1}-\Omega,\Omega-\omega_{3}},
\nonumber
\end{align}
and
\begin{align}
&
 iS_{\mathrm{interaction};4}=
-i\frac{J}{4}
\int_{\{\omega\}}
\int_{\{{\bf k}\}}
\gamma_{{\bf k}_{2}}^{*}
 b_{{\bf k}_{1}}^{\dag}a_{{\bf k}_{2}}b_{{\bf k}_{3}}^{\dag}b_{{\bf k}_{4}}
\delta_{{\bf k}_{1}-{\bf k}_{2},{\bf k}_{4}-{\bf k}_{3}} 
\delta_{\omega_{1}-\omega_{2},\omega_{4}-\omega_{3}}
\\
&
\rightarrow
-i
\frac{J}{8}
\int_{\{\omega\}}
\int_{\{{\bf k}\}}
\gamma_{{\bf k}_{2}}^{*}
\bigg(
\bar{\Psi}^{\mathrm{cl}}_{\beta;{\bf k}_{1};\omega_{1}} 
\bar{\Psi}^{\mathrm{cl}}_{\beta;{\bf k}_{3};\omega_{3}}
\Psi^{\mathrm{cl}}_{\alpha;{\bf k}_{2};\omega_{2}}
\Psi^{\mathrm{q}}_{\beta;{\bf k}_{4};\omega_{4}} 
+
\bar{\Psi}^{\mathrm{cl}}_{\beta;{\bf k}_{1};\omega_{1}} 
\bar{\Psi}^{\mathrm{q}}_{\beta;{\bf k}_{3};\omega_{3}}
\Psi^{\mathrm{cl}}_{\alpha;{\bf k}_{2};\omega_{2}}
\Psi^{\mathrm{cl}}_{\beta;{\bf k}_{4};\omega_{4}} 
\nonumber
\\
&
+
\bar{\Psi}^{\mathrm{cl}}_{\beta;{\bf k}_{1};\omega_{1}} 
\bar{\Psi}^{\mathrm{cl}}_{\beta;{\bf k}_{3};\omega_{3}}
\Psi^{\mathrm{q}}_{\alpha;{\bf k}_{2};\omega_{2}}
\Psi^{\mathrm{cl}}_{\beta;{\bf k}_{4};\omega_{4}} 
+
\bar{\Psi}^{\mathrm{q}}_{\beta;{\bf k}_{1};\omega_{1}} 
\bar{\Psi}^{\mathrm{cl}}_{\beta;{\bf k}_{3};\omega_{3}}
\Psi^{\mathrm{cl}}_{\alpha;{\bf k}_{2};\omega_{2}}
\Psi^{\mathrm{cl}}_{\beta;{\bf k}_{4};\omega_{4}} 
\bigg)
\delta_{{\bf k}_{1}-{\bf k}_{2},{\bf k}_{4}-{\bf k}_{3}} 
\delta_{\omega_{1}-\omega_{2},\omega_{4}-\omega_{3}}
\nonumber
\\
&
-i
\frac{J}{8}
\int_{\{\omega\}}
\int_{\{{\bf k}\}}
\gamma_{{\bf k}_{2}}^{*}
\left(\frac{\Gamma \sqrt{S}}{3SJ}\right)^2
\Psi^{\mathrm{cl}}_{\alpha;{\bf k}_{2};\omega_{2}}
\Psi^{\mathrm{q}}_{\beta;{\bf k}_{4};\omega_{4}} 
\delta_{-{\bf k}_{2},{\bf k}_{4}} 
\delta_{\Omega-\omega_{2},\omega_{4}-\Omega}
\\
&
-i
\frac{J}{8}
\int_{\{\omega\}}
\int_{\{{\bf k}\}}
\gamma_{0}
\left(\frac{\Gamma \sqrt{S}}{3SJ}\right)^2
\bar{\Psi}^{\mathrm{cl}}_{\beta;{\bf k}_{1};\omega_{1}} 
\bar{\Psi}^{\mathrm{q}}_{\beta;{\bf k}_{3};\omega_{3}}
\delta_{{\bf k}_{1},-{\bf k}_{3}}
\delta_{\omega_{1}-\Omega,\Omega-\omega_{3}}
\nonumber
\\
&
-i
\frac{J}{8}
\int_{\{\omega\}}
\int_{\{{\bf k}\}}
\gamma_{{\bf k}_{2}}^{*}
\left(\frac{\Gamma \sqrt{S}}{3SJ}\right)^2
\Psi^{\mathrm{q}}_{\alpha;{\bf k}_{2};\omega_{2}}
\Psi^{\mathrm{cl}}_{\beta;{\bf k}_{4};\omega_{4}} 
\delta_{-{\bf k}_{2},{\bf k}_{4}} 
\delta_{\Omega-\omega_{2},\omega_{4}-\Omega}
\\
&
-i
\frac{J}{8}
\int_{\{\omega\}}
\int_{\{{\bf k}\}}
\gamma_{0}
\left(\frac{\Gamma \sqrt{S}}{3SJ}\right)^2
\bar{\Psi}^{\mathrm{q}}_{\beta;{\bf k}_{1};\omega_{1}} 
\bar{\Psi}^{\mathrm{cl}}_{\beta;{\bf k}_{3};\omega_{3}}
\delta_{{\bf k}_{1},-{\bf k}_{3}}
\delta_{\omega_{1}-\Omega,\Omega-\omega_{3}}.
\nonumber
\end{align}
There is also $\propto -J$ interaction term, which also gets shifted accordingly. 
\begin{align}
&
 iS_{\mathrm{interaction};5}=
i J
\int_{\{\omega\}}
\int_{\{{\bf k}\}}
\gamma_{{\bf k}_{4}-{\bf k}_{3}}
 a_{{\bf k}_{1}}^{\dag}a_{{\bf k}_{2}}
 b_{{\bf k}_{3}}^{\dag}b_{{\bf k}_{4}}
\delta_{{\bf k}_{1}-{\bf k}_{2},{\bf k}_{4}-{\bf k}_{3}} 
\delta_{\omega_{1}-\omega_{2},\omega_{4}-\omega_{3}}
\\
&
\rightarrow
iJ\frac{1}{2}
\int_{\{\omega\}}
\int_{\{{\bf k}\}}
\gamma_{{\bf k}_{4}-{\bf k}_{3}}
\bigg(
\bar{\Psi}^{\mathrm{cl}}_{\alpha;{\bf k}_{1};\omega_{1}} 
\bar{\Psi}^{\mathrm{cl}}_{\beta;{\bf k}_{3};\omega_{3}}
\Psi^{\mathrm{cl}}_{\alpha;{\bf k}_{2};\omega_{2}}
\Psi^{\mathrm{q}}_{\beta;{\bf k}_{4};\omega_{4}} 
+
\bar{\Psi}^{\mathrm{cl}}_{\alpha;{\bf k}_{1};\omega_{1}} 
\bar{\Psi}^{\mathrm{q}}_{\beta;{\bf k}_{3};\omega_{3}}
\Psi^{\mathrm{cl}}_{\alpha;{\bf k}_{2};\omega_{2}}
\Psi^{\mathrm{cl}}_{\beta;{\bf k}_{4};\omega_{4}} 
\nonumber
\\
&
+
\bar{\Psi}^{\mathrm{cl}}_{\alpha;{\bf k}_{1};\omega_{1}} 
\bar{\Psi}^{\mathrm{cl}}_{\beta;{\bf k}_{3};\omega_{3}}
\Psi^{\mathrm{q}}_{\alpha;{\bf k}_{2};\omega_{2}}
\Psi^{\mathrm{cl}}_{\beta;{\bf k}_{4};\omega_{4}} 
+
\bar{\Psi}^{\mathrm{q}}_{\alpha;{\bf k}_{1};\omega_{1}} 
\bar{\Psi}^{\mathrm{cl}}_{\beta;{\bf k}_{3};\omega_{3}}
\Psi^{\mathrm{cl}}_{\alpha;{\bf k}_{2};\omega_{2}}
\Psi^{\mathrm{cl}}_{\beta;{\bf k}_{4};\omega_{4}} 
\bigg)
\delta_{{\bf k}_{1}-{\bf k}_{2},{\bf k}_{4}-{\bf k}_{3}} 
\delta_{\omega_{1}-\omega_{2},\omega_{4}-\omega_{3}}
\nonumber
\\
&
+
i J
\frac{1}{2}
\int_{\{\omega\}}
\int_{\{{\bf k}\}}
\gamma_{{\bf k}_{4}}
\left(\frac{\Gamma \sqrt{S}}{3SJ}\right)^2
\Psi^{\mathrm{cl}}_{\alpha;{\bf k}_{2};\omega_{2}}
\Psi^{\mathrm{q}}_{\beta;{\bf k}_{4};\omega_{4}} 
\delta_{-{\bf k}_{2},{\bf k}_{4}} 
\delta_{\Omega-\omega_{2},\omega_{4}-\Omega}
\\
&
+
i J
\frac{1}{2}
\int_{\{\omega\}}
\int_{\{{\bf k}\}}
\gamma_{-{\bf k}_{3}}
\left(\frac{\Gamma \sqrt{S}}{3SJ}\right)^2
\bar{\Psi}^{\mathrm{cl}}_{\alpha;{\bf k}_{1};\omega_{1}} 
\bar{\Psi}^{\mathrm{q}}_{\beta;{\bf k}_{3};\omega_{3}}
\delta_{{\bf k}_{1},-{\bf k}_{3}}
\delta_{\omega_{1}-\Omega,\Omega-\omega_{3}}
\nonumber
\\
&
+
 iJ
\frac{1}{2}
\int_{\{\omega\}}
\int_{\{{\bf k}\}}
\gamma_{{\bf k}_{4}}
\left(\frac{\Gamma \sqrt{S}}{3SJ}\right)^2
\Psi^{\mathrm{q}}_{\alpha;{\bf k}_{2};\omega_{2}}
\Psi^{\mathrm{cl}}_{\beta;{\bf k}_{4};\omega_{4}} 
\delta_{-{\bf k}_{2},{\bf k}_{4}} 
\delta_{\Omega-\omega_{2},\omega_{4}-\Omega}
\\
&
+
iJ
\frac{1}{2}
\int_{\{\omega\}}
\int_{\{{\bf k}\}}
\gamma_{-{\bf k}_{3}}
\left(\frac{\Gamma \sqrt{S}}{3SJ}\right)^2
\bar{\Psi}^{\mathrm{q}}_{\alpha;{\bf k}_{1};\omega_{1}} 
\bar{\Psi}^{\mathrm{cl}}_{\beta;{\bf k}_{3};\omega_{3}}
\delta_{{\bf k}_{1},-{\bf k}_{3}}
\delta_{\omega_{1}-\Omega,\Omega-\omega_{3}}.
\end{align}
Collecting now terms quadratic in fields, we get for the pump 
\begin{align}
&H_{\mathrm{pump}} 
\\
&
= 
\frac{J}{4}
\left(\frac{\Gamma \sqrt{S}}{3SJ}\right)^2
\int_{\{{\bf k}\}}
\bigg[
-
\gamma_{{\bf k}_{4}}
\Psi^{\mathrm{cl}}_{\alpha;{\bf k}_{2};\omega_{2}}
\Psi^{\mathrm{q}}_{\beta;{\bf k}_{4};\omega_{4}} 
\delta_{-{\bf k}_{2},{\bf k}_{4}} 
\delta_{\Omega-\omega_{2},\omega_{4}-\Omega}
-
\gamma_{{\bf k}_{4}}
\Psi^{\mathrm{q}}_{\alpha;{\bf k}_{2};\omega_{2}}
\Psi^{\mathrm{cl}}_{\beta;{\bf k}_{4};\omega_{4}} 
\delta_{-{\bf k}_{2},{\bf k}_{4}} 
\delta_{\Omega-\omega_{2},\omega_{4}-\Omega}
\nonumber
\\
&
-
\gamma_{{\bf k}_{3}}^{*} 
\bar{\Psi}^{\mathrm{cl}}_{\alpha;{\bf k}_{1};\omega_{1}} 
\bar{\Psi}^{\mathrm{q}}_{\beta;{\bf k}_{3};\omega_{3}}
\delta_{{\bf k}_{1},-{\bf k}_{3}} 
\delta_{\omega_{1}-\Omega,\Omega-\omega_{3}}
-
\gamma_{{\bf k}_{3}}^{*} 
\bar{\Psi}^{\mathrm{q}}_{\alpha;{\bf k}_{1};\omega_{1}} 
\bar{\Psi}^{\mathrm{cl}}_{\beta;{\bf k}_{3};\omega_{3}}
\delta_{{\bf k}_{1},-{\bf k}_{3}} 
\delta_{\omega_{1}-\Omega,\Omega-\omega_{3}}
\nonumber
\\
&
+
\gamma_{0} 
\Psi^{\mathrm{cl}}_{\alpha;{\bf k}_{2};\omega_{2}}
\Psi^{\mathrm{q}}_{\alpha;{\bf k}_{4};\omega_{4}} 
\delta_{-{\bf k}_{2},{\bf k}_{4}} 
\delta_{\Omega-\omega_{2},\omega_{4}-\Omega}
+
\gamma_{0}
\bar{\Psi}^{\mathrm{cl}}_{\alpha;{\bf k}_{1};\omega_{1}} 
\bar{\Psi}^{\mathrm{q}}_{\alpha;{\bf k}_{3};\omega_{3}}
\delta_{{\bf k}_{1},-{\bf k}_{3}}
\delta_{\omega_{1}-\Omega,\Omega-\omega_{3}}
\nonumber
\\
&
+
\gamma_{0} 
\Psi^{\mathrm{cl}}_{\beta;{\bf k}_{2};\omega_{2}}
\Psi^{\mathrm{q}}_{\beta;{\bf k}_{4};\omega_{4}} 
\delta_{-{\bf k}_{2},{\bf k}_{4}} 
\delta_{\Omega-\omega_{2},\omega_{4}-\Omega}
+
\gamma_{0}
\bar{\Psi}^{\mathrm{cl}}_{\beta;{\bf k}_{1};\omega_{1}} 
\bar{\Psi}^{\mathrm{q}}_{\beta;{\bf k}_{3};\omega_{3}}
\delta_{{\bf k}_{1},-{\bf k}_{3}}
\delta_{\omega_{1}-\Omega,\Omega-\omega_{3}}
\nonumber
\bigg].
\end{align}

Let us now demonstrate that indeed terms of the $\propto \bar{\Psi}^{\mathrm{q}}_{\alpha;{\bf k}_{1};\omega_{1}} 
\Psi^{\mathrm{cl}}_{\alpha;{\bf k}_{2};\omega_{2}}$ type sum up to zero and, hence, can't be generated by the pump process. Recall, that overall there are five interaction terms listed in this subsection. We refer to them in the order they have appeared. From the first interaction term we have
\begin{align}
&
\frac{J}{8}
\int_{\{\omega\}}
\int_{\{{\bf k}\}}
\gamma_{{\bf k}_{4}}
\bigg(
\bar{\Psi}^{\mathrm{cl}}_{\alpha;{\bf k}_{1};\omega_{1}} 
\Psi^{\mathrm{cl}}_{\alpha;{\bf k}_{2};\omega_{2}}
\bar{\Psi}^{\mathrm{cl}}_{\alpha;{\bf k}_{3};\omega_{3}}
\Psi^{\mathrm{q}}_{\beta;{\bf k}_{4};\omega_{4}} 
+
\bar{\Psi}^{\mathrm{cl}}_{\alpha;{\bf k}_{1};\omega_{1}} 
\Psi^{\mathrm{cl}}_{\alpha;{\bf k}_{2};\omega_{2}}
\bar{\Psi}^{\mathrm{q}}_{\alpha;{\bf k}_{3};\omega_{3}}
\Psi^{\mathrm{cl}}_{\beta;{\bf k}_{4};\omega_{4}} 
\nonumber
\\
&
+
\bar{\Psi}^{\mathrm{cl}}_{\alpha;{\bf k}_{1};\omega_{1}} 
\Psi^{\mathrm{q}}_{\alpha;{\bf k}_{2};\omega_{2}}
\bar{\Psi}^{\mathrm{cl}}_{\alpha;{\bf k}_{3};\omega_{3}}
\Psi^{\mathrm{cl}}_{\beta;{\bf k}_{4};\omega_{4}} 
+
\bar{\Psi}^{\mathrm{q}}_{\alpha;{\bf k}_{1};\omega_{1}} 
\Psi^{\mathrm{cl}}_{\alpha;{\bf k}_{2};\omega_{2}}
\bar{\Psi}^{\mathrm{cl}}_{\alpha;{\bf k}_{3};\omega_{3}}
\Psi^{\mathrm{cl}}_{\beta;{\bf k}_{4};\omega_{4}} 
\bigg)
\delta_{{\bf k}_{1}-{\bf k}_{2},{\bf k}_{4}-{\bf k}_{3}} 
\delta_{\omega_{1}-\omega_{2},\omega_{4}-\omega_{3}}
\nonumber
\\
&
\rightarrow
\frac{J}{4}
\int_{\{\omega\}}
\int_{\{{\bf k}\}}
\gamma_{0}
\bar{\Psi}^{\mathrm{q}}_{\alpha;{\bf k}_{1};\omega_{1}} 
\Psi^{\mathrm{cl}}_{\alpha;{\bf k}_{2};\omega_{2}}
\delta_{{\bf k}_{1},{\bf k}_{2}}\delta_{\omega_{1},\omega_{2}}.
\end{align}
From the second interaction term we have
\begin{align}
&
\frac{J}{8}
\int_{\{\omega\}}
\int_{\{{\bf k}\}}
\gamma_{{\bf k}_{3}}^{*} 
\bigg(
\bar{\Psi}^{\mathrm{cl}}_{\alpha;{\bf k}_{1};\omega_{1}} 
\Psi^{\mathrm{cl}}_{\alpha;{\bf k}_{2};\omega_{2}}
\bar{\Psi}^{\mathrm{cl}}_{\beta;{\bf k}_{3};\omega_{3}}
\Psi^{\mathrm{q}}_{\alpha;{\bf k}_{4};\omega_{4}} 
+
\bar{\Psi}^{\mathrm{cl}}_{\alpha;{\bf k}_{1};\omega_{1}} 
\Psi^{\mathrm{cl}}_{\alpha;{\bf k}_{2};\omega_{2}}
\bar{\Psi}^{\mathrm{q}}_{\beta;{\bf k}_{3};\omega_{3}}
\Psi^{\mathrm{cl}}_{\alpha;{\bf k}_{4};\omega_{4}} 
\nonumber
\\
&
+
\bar{\Psi}^{\mathrm{cl}}_{\alpha;{\bf k}_{1};\omega_{1}} 
\Psi^{\mathrm{q}}_{\alpha;{\bf k}_{2};\omega_{2}}
\bar{\Psi}^{\mathrm{cl}}_{\beta;{\bf k}_{3};\omega_{3}}
\Psi^{\mathrm{cl}}_{\alpha;{\bf k}_{4};\omega_{4}} 
+
\bar{\Psi}^{\mathrm{q}}_{\alpha;{\bf k}_{1};\omega_{1}} 
\Psi^{\mathrm{cl}}_{\alpha;{\bf k}_{2};\omega_{2}}
\bar{\Psi}^{\mathrm{cl}}_{\beta;{\bf k}_{3};\omega_{3}}
\Psi^{\mathrm{cl}}_{\alpha;{\bf k}_{4};\omega_{4}} 
\bigg)
\delta_{{\bf k}_{1}-{\bf k}_{2},{\bf k}_{4}-{\bf k}_{3}} 
\delta_{\omega_{1}-\omega_{2},\omega_{4}-\omega_{3}}
\nonumber
\\
&
\rightarrow
\frac{J}{4}
\int_{\{\omega\}}
\int_{\{{\bf k}\}}
\gamma_{0}
\bar{\Psi}^{\mathrm{q}}_{\alpha;{\bf k}_{1};\omega_{1}} 
\Psi^{\mathrm{cl}}_{\alpha;{\bf k}_{2};\omega_{2}}
\delta_{{\bf k}_{1},{\bf k}_{2}}\delta_{\omega_{1},\omega_{2}}.
\end{align}
From the fifth interaction term we have
\begin{align}
&
-J\frac{1}{2}
\int_{\{\omega\}}
\int_{\{{\bf k}\}}
\gamma_{{\bf k}_{4}-{\bf k}_{3}}
\bigg(
\bar{\Psi}^{\mathrm{cl}}_{\alpha;{\bf k}_{1};\omega_{1}} 
\Psi^{\mathrm{cl}}_{\alpha;{\bf k}_{2};\omega_{2}}
\bar{\Psi}^{\mathrm{cl}}_{\beta;{\bf k}_{3};\omega_{3}}
\Psi^{\mathrm{q}}_{\beta;{\bf k}_{4};\omega_{4}} 
+
\bar{\Psi}^{\mathrm{cl}}_{\alpha;{\bf k}_{1};\omega_{1}} 
\Psi^{\mathrm{cl}}_{\alpha;{\bf k}_{2};\omega_{2}}
\bar{\Psi}^{\mathrm{q}}_{\beta;{\bf k}_{3};\omega_{3}}
\Psi^{\mathrm{cl}}_{\beta;{\bf k}_{4};\omega_{4}} 
\nonumber
\\
&
+
\bar{\Psi}^{\mathrm{cl}}_{\alpha;{\bf k}_{1};\omega_{1}} 
\Psi^{\mathrm{q}}_{\alpha;{\bf k}_{2};\omega_{2}}
\bar{\Psi}^{\mathrm{cl}}_{\beta;{\bf k}_{3};\omega_{3}}
\Psi^{\mathrm{cl}}_{\beta;{\bf k}_{4};\omega_{4}} 
+
\bar{\Psi}^{\mathrm{q}}_{\alpha;{\bf k}_{1};\omega_{1}} 
\Psi^{\mathrm{cl}}_{\alpha;{\bf k}_{2};\omega_{2}}
\bar{\Psi}^{\mathrm{cl}}_{\beta;{\bf k}_{3};\omega_{3}}
\Psi^{\mathrm{cl}}_{\beta;{\bf k}_{4};\omega_{4}} 
\bigg)
\delta_{{\bf k}_{1}-{\bf k}_{2},{\bf k}_{4}-{\bf k}_{3}} 
\delta_{\omega_{1}-\omega_{2},\omega_{4}-\omega_{3}}
\nonumber
\\
&
\rightarrow 
-\frac{J}{2}
\int_{\{\omega\}}
\int_{\{{\bf k}\}}
\gamma_{0}
\bar{\Psi}^{\mathrm{q}}_{\alpha;{\bf k}_{1};\omega_{1}} 
\Psi^{\mathrm{cl}}_{\alpha;{\bf k}_{2};\omega_{2}} \delta_{{\bf k}_{1},{\bf k}_{2}}\delta_{\omega_{1},\omega_{2}}.
\end{align}
Three terms sum up to zero. The same can be proven for the other combinations of the same type.

\subsection{Hartree-Fock corrections}
\begin{figure}[h] 
\centerline{
\includegraphics[width=0.25\textwidth,height=0.08\textheight]{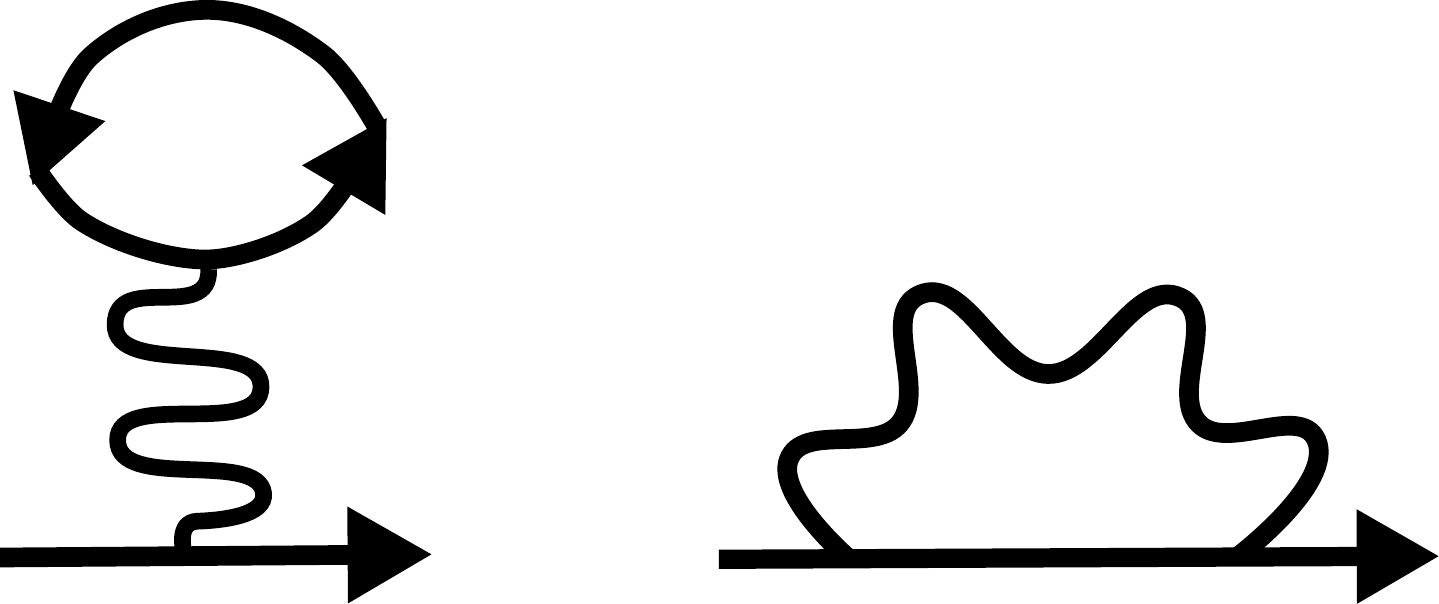}
}
\protect\caption{Hartree-Fock corrections to the magnon dispersion. }

\label{fig:HartreeFock}  

\end{figure}
In order to understand possible instabilities in the system due to the magnon pair creation, we also need to take in to account Hartree-Fock corrections to the magnon dispersion \cite{BlochPRL1962,Pershoguba}. They are expected to give temperature dependent correction, and, thus, might be important when discussing the experimental details. 
For example, let us pick the first interaction term,
\begin{align}\label{firstHF}
\langle iS_{\mathrm{interaction};1}\rangle
=
&
-
i\frac{J}{8}
\int_{\{\omega\}}
\int_{\{{\bf k}\}}
\gamma_{{\bf k}_{4}}
\langle
\bar{\Psi}^{\mathrm{cl}}_{\alpha;{\bf k}_{1};\omega_{1}} 
\bar{\Psi}^{\mathrm{cl}}_{\alpha;{\bf k}_{3};\omega_{3}}
\Psi^{\mathrm{cl}}_{\alpha;{\bf k}_{2};\omega_{2}}
\Psi^{\mathrm{q}}_{\beta;{\bf k}_{4};\omega_{4}} 
+
\bar{\Psi}^{\mathrm{cl}}_{\alpha;{\bf k}_{1};\omega_{1}} 
\bar{\Psi}^{\mathrm{q}}_{\alpha;{\bf k}_{3};\omega_{3}}
\Psi^{\mathrm{cl}}_{\alpha;{\bf k}_{2};\omega_{2}}
\Psi^{\mathrm{cl}}_{\beta;{\bf k}_{4};\omega_{4}} 
\nonumber
\\
&
+
\bar{\Psi}^{\mathrm{cl}}_{\alpha;{\bf k}_{1};\omega_{1}} 
\bar{\Psi}^{\mathrm{cl}}_{\alpha;{\bf k}_{3};\omega_{3}}
\Psi^{\mathrm{q}}_{\alpha;{\bf k}_{2};\omega_{2}}
\Psi^{\mathrm{cl}}_{\beta;{\bf k}_{4};\omega_{4}} 
+
\bar{\Psi}^{\mathrm{q}}_{\alpha;{\bf k}_{1};\omega_{1}} 
\bar{\Psi}^{\mathrm{cl}}_{\alpha;{\bf k}_{3};\omega_{3}}
\Psi^{\mathrm{cl}}_{\alpha;{\bf k}_{2};\omega_{2}}
\Psi^{\mathrm{cl}}_{\beta;{\bf k}_{4};\omega_{4}} 
\rangle
\delta_{{\bf k}_{1}-{\bf k}_{2},{\bf k}_{4}-{\bf k}_{3}} 
\delta_{\omega_{1}-\omega_{2},\omega_{4}-\omega_{3}}.
\end{align}
We found that for the task at hand it is more convenient to come back to time domain rather to work in frequency domain.
In this way, equal-time commutation relations 
\begin{align}
[\Psi^{\mathrm{cl}}_{n;{\bf k}_{1}}(t),\bar{\Psi}^{\mathrm{cl}}_{m;{\bf k}_{2}}(t)] = 
\delta_{n,m}
\delta_{{\bf k}_{1},{\bf k}_{2}}
\end{align}
are written in the most transparent way.
For example, picking the first term in Eq. (\ref{firstHF}),
\begin{align}
&
\frac{J}{8}
\int_{\{\omega\}}
\int_{\{{\bf k}\}}
\gamma_{{\bf k}_{4}}
\langle
\bar{\Psi}^{\mathrm{cl}}_{\alpha;{\bf k}_{1};\omega_{1}} 
\bar{\Psi}^{\mathrm{cl}}_{\alpha;{\bf k}_{3};\omega_{3}}
\Psi^{\mathrm{cl}}_{\alpha;{\bf k}_{2};\omega_{2}}
\Psi^{\mathrm{q}}_{\beta;{\bf k}_{4};\omega_{4}} 
\rangle
\delta_{{\bf k}_{1}-{\bf k}_{2},{\bf k}_{4}-{\bf k}_{3}} 
\delta_{\omega_{1}-\omega_{2},\omega_{4}-\omega_{3}}
\\
=
&
\frac{J}{8}
\int_{t}
\int_{\{{\bf k}\}}
\gamma_{{\bf k}_{4}}
\bar{\Psi}^{\mathrm{cl}}_{\alpha;{\bf k}_{1}}(t) 
\Psi^{\mathrm{q}}_{\beta;{\bf k}_{4}}(t) 
\langle
\bar{\Psi}^{\mathrm{cl}}_{\alpha;{\bf k}_{3}}(t)
\Psi^{\mathrm{cl}}_{\alpha;{\bf k}_{2}}(t)
\rangle
\delta_{{\bf k}_{1}-{\bf k}_{2},{\bf k}_{4}-{\bf k}_{3}} 
\\
+
&
\int_{t}
\int_{\{{\bf k}\}}
\gamma_{{\bf k}_{4}}
\bar{\Psi}^{\mathrm{cl}}_{\alpha;{\bf k}_{3}}(t)
\Psi^{\mathrm{q}}_{\beta;{\bf k}_{4}}(t)
\langle
\bar{\Psi}^{\mathrm{cl}}_{\alpha;{\bf k}_{1}}(t)
\Psi^{\mathrm{cl}}_{\alpha;{\bf k}_{2}}(t)
\rangle
\delta_{{\bf k}_{1}-{\bf k}_{2},{\bf k}_{4}-{\bf k}_{3}} 
\\
=
&
\frac{J}{4}
\int_{t}
\int_{{\bf k}}
\gamma_{{\bf k}}
\bar{\Psi}^{\mathrm{cl}}_{\alpha;{\bf k}}(t)
\Psi^{\mathrm{q}}_{\beta;{\bf k}}(t)
\int_{\bf q}
\left[ -1 + i\int_{\epsilon} G^{\mathrm{K}}_{\alpha\alpha}(\epsilon;{\bf q})\right]
\\
=
&
\frac{J}{4}
\int_{t}
\int_{{\bf k}}
\gamma_{{\bf k}}
\bar{\Psi}^{\mathrm{cl}}_{\alpha;{\bf k}}(t)
\Psi^{\mathrm{q}}_{\beta;{\bf k}}(t)
\int_{\bf q}
\left[ n_{\mathrm{B}}(\epsilon_{+;{\bf q}})+ n_{\mathrm{B}}(\epsilon_{-;{\bf q}})\right],
\end{align}
where we used ${\cal F}(\epsilon) = 1 + \frac{2}{e^{\frac{\epsilon}{T}}-1} \equiv 1+2n_{\mathrm{B}}(\epsilon)$ identity, and where $-1$ in the $\left[ -1 + i\int_{\epsilon} G^{\mathrm{K}}_{\alpha\alpha}(\epsilon;{\bf q})\right]$ factor is due to the commutation relations.
Now picking the second term in Eq. (\ref{firstHF}),
\begin{align}
&
\frac{J}{8}
\int_{\{\omega\}}
\int_{\{{\bf k}\}}
\gamma_{{\bf k}_{4}}
\langle
\bar{\Psi}^{\mathrm{cl}}_{\alpha;{\bf k}_{1};\omega_{1}} 
\bar{\Psi}^{\mathrm{q}}_{\alpha;{\bf k}_{3};\omega_{3}}
\Psi^{\mathrm{cl}}_{\alpha;{\bf k}_{2};\omega_{2}}
\Psi^{\mathrm{cl}}_{\beta;{\bf k}_{4};\omega_{4}} 
\rangle
\delta_{{\bf k}_{1}-{\bf k}_{2},{\bf k}_{4}-{\bf k}_{3}} 
\delta_{\omega_{1}-\omega_{2},\omega_{4}-\omega_{3}}
\\
=
&
\frac{J}{8}
\int_{t}
\int_{\bf k}
\bar{\Psi}^{\mathrm{q}}_{\alpha;{\bf k}}(t)
\Psi^{\mathrm{cl}}_{\alpha;{\bf k}}(t)
\int_{\bf q}
\int_{\epsilon}
iG^{\mathrm{K}}_{\beta\alpha}(\epsilon;{\bf q})
\gamma_{\bf q}
+
\frac{J}{8}
\int_{t}
\int_{{\bf k}}
\gamma_{{\bf k}}
\bar{\Psi}^{\mathrm{q}}_{\alpha;{\bf k}}(t)
\Psi^{\mathrm{cl}}_{\beta;{\bf k}}(t)
\int_{\bf q}
\left[ -1 + i\int_{\epsilon} G^{\mathrm{K}}_{\alpha\alpha}(\epsilon;{\bf q})\right]
\\
=
&
-
\frac{J}{8}
\int_{t}
\int_{\bf k}
\bar{\Psi}^{\mathrm{q}}_{\alpha;{\bf k}}(t)
\Psi^{\mathrm{cl}}_{\alpha;{\bf k}}(t)
\int_{\bf q}
\vert \gamma_{\bf q} \vert
\left[ n_{\mathrm{B}}(\epsilon_{+;{\bf q}}) - n_{\mathrm{B}}(\epsilon_{-;{\bf q}})\right]
+
\frac{J}{8}
\int_{t}
\int_{{\bf k}}
\gamma_{{\bf k}}
\bar{\Psi}^{\mathrm{q}}_{\alpha;{\bf k}}(t)
\Psi^{\mathrm{cl}}_{\beta;{\bf k}}(t)
\int_{\bf q}
\left[ n_{\mathrm{B}}(\epsilon_{+;{\bf q}}) + n_{\mathrm{B}}(\epsilon_{-;{\bf q}})\right].
\end{align}
Third term in Eq. (\ref{firstHF}) reads,
\begin{align}
&
\frac{J}{8}
\int_{\{\omega\}}
\int_{\{{\bf k}\}}
\gamma_{{\bf k}_{4}}
\langle
\bar{\Psi}^{\mathrm{cl}}_{\alpha;{\bf k}_{1};\omega_{1}} 
\bar{\Psi}^{\mathrm{cl}}_{\alpha;{\bf k}_{3};\omega_{3}}
\Psi^{\mathrm{q}}_{\alpha;{\bf k}_{2};\omega_{2}}
\Psi^{\mathrm{cl}}_{\beta;{\bf k}_{4};\omega_{4}} 
\rangle
\delta_{{\bf k}_{1}-{\bf k}_{2},{\bf k}_{4}-{\bf k}_{3}} 
\delta_{\omega_{1}-\omega_{2},\omega_{4}-\omega_{3}}
\\
=
&
\frac{J}{4}
\int_{t}
\int_{\bf k}
\bar{\Psi}^{\mathrm{cl}}_{\alpha;{\bf k}}(t)
\Psi^{\mathrm{q}}_{\alpha;{\bf k}}(t)
\int_{\bf q}
\int_{\epsilon}
iG^{\mathrm{K}}_{\beta\alpha}(\epsilon;{\bf q})
\gamma_{\bf q}
\\
=
&
-
\frac{J}{4}
\int_{t}
\int_{\bf k}
\bar{\Psi}^{\mathrm{cl}}_{\alpha;{\bf k}}(t)
\Psi^{\mathrm{q}}_{\alpha;{\bf k}}(t)
\int_{\bf q}
\vert \gamma_{\bf q} \vert
\left[ n_{\mathrm{B}}(\epsilon_{+;{\bf q}}) - n_{\mathrm{B}}(\epsilon_{-;{\bf q}})\right].
\end{align}
Finally, the last term in Eq. (\ref{firstHF}) reads,
\begin{align}
&
\frac{J}{8}
\int_{\{\omega\}}
\int_{\{{\bf k}\}}
\gamma_{{\bf k}_{4}}
\langle
\bar{\Psi}^{\mathrm{q}}_{\alpha;{\bf k}_{1};\omega_{1}} 
\bar{\Psi}^{\mathrm{cl}}_{\alpha;{\bf k}_{3};\omega_{3}}
\Psi^{\mathrm{cl}}_{\alpha;{\bf k}_{2};\omega_{2}}
\Psi^{\mathrm{cl}}_{\beta;{\bf k}_{4};\omega_{4}} 
\rangle
\delta_{{\bf k}_{1}-{\bf k}_{2},{\bf k}_{4}-{\bf k}_{3}} 
\delta_{\omega_{1}-\omega_{2},\omega_{4}-\omega_{3}}
\\
=
&
-
\frac{J}{8}
\int_{t}
\int_{\bf k}
\bar{\Psi}^{\mathrm{q}}_{\alpha;{\bf k}}(t)
\Psi^{\mathrm{cl}}_{\alpha;{\bf k}}(t)
\int_{\bf q}
\vert \gamma_{\bf q} \vert
\left[ n_{\mathrm{B}}(\epsilon_{+;{\bf q}}) - n_{\mathrm{B}}(\epsilon_{-;{\bf q}})\right]
+
\frac{J}{8}
\int_{t}
\int_{{\bf k}}
\gamma_{{\bf k}}
\bar{\Psi}^{\mathrm{q}}_{\alpha;{\bf k}}(t)
\Psi^{\mathrm{cl}}_{\beta;{\bf k}}(t)
\int_{\bf q}
\left[ n_{\mathrm{B}}(\epsilon_{+;{\bf q}}) + n_{\mathrm{B}}(\epsilon_{-;{\bf q}})\right],
\end{align}
which essentially doubles the second term in Eq. (\ref{firstHF}). Collecting all the four terms, we get
\begin{align}
\langle iS_{\mathrm{interaction};1}\rangle = & 
-i
\frac{J}{4}
I_{1}
\int_{t}
\int_{{\bf k}}
\gamma_{{\bf k}}
\bar{\Psi}^{\mathrm{cl}}_{\alpha;{\bf k}}(t)
\Psi^{\mathrm{q}}_{\beta;{\bf k}}(t)
-i
\frac{J}{4}
I_{2}
\int_{t}
\int_{\bf k}
\bar{\Psi}^{\mathrm{cl}}_{\alpha;{\bf k}}(t)
\Psi^{\mathrm{q}}_{\alpha;{\bf k}}(t)
\\
&
-i
\frac{J}{4}
I_{2}
\int_{t}
\int_{\bf k}
\bar{\Psi}^{\mathrm{q}}_{\alpha;{\bf k}}(t)
\Psi^{\mathrm{cl}}_{\alpha;{\bf k}}(t)
-i
\frac{J}{4}
I_{1}
\int_{t}
\int_{{\bf k}}
\gamma_{{\bf k}}
\bar{\Psi}^{\mathrm{q}}_{\alpha;{\bf k}}(t)
\Psi^{\mathrm{cl}}_{\beta;{\bf k}}(t),
\end{align}
where 
\begin{align}
&
I_{1} = 
\int_{\bf q}
\left[ -1 + i\int_{\epsilon} G^{\mathrm{K}}_{\alpha\alpha}(\epsilon;{\bf q})\right]
=
\int_{\bf q}
\left[ n_{\mathrm{B}}(\epsilon_{+;{\bf q}}) + n_{\mathrm{B}}(\epsilon_{-;{\bf q}})\right],
\\
&
I_{2} =
\int_{\bf q}
\int_{\epsilon}
iG^{\mathrm{K}}_{\beta\alpha}(\epsilon;{\bf q})
\gamma_{\bf q}
=- \int_{\bf q}
\vert \gamma_{\bf q} \vert
\left[ n_{\mathrm{B}}(\epsilon_{+;{\bf q}}) - n_{\mathrm{B}}(\epsilon_{-;{\bf q}})\right].
\end{align}
Expressions for the three other interaction terms, i.e. $\langle iS_{\mathrm{interaction};2,3,4}\rangle $, are similar to the obtained one. Fifth interaction is
\begin{align}
\langle iS_{\mathrm{interaction};5}\rangle =
&
i\frac{J}{2}
\int_{\{\omega\}}
\int_{\{{\bf k}\}}
\gamma_{{\bf k}_{4}-{\bf k}_{3}}
\langle
\bar{\Psi}^{\mathrm{cl}}_{\alpha;{\bf k}_{1};\omega_{1}} 
\bar{\Psi}^{\mathrm{cl}}_{\beta;{\bf k}_{3};\omega_{3}}
\Psi^{\mathrm{cl}}_{\alpha;{\bf k}_{2};\omega_{2}}
\Psi^{\mathrm{q}}_{\beta;{\bf k}_{4};\omega_{4}} 
+
\bar{\Psi}^{\mathrm{cl}}_{\alpha;{\bf k}_{1};\omega_{1}} 
\bar{\Psi}^{\mathrm{q}}_{\beta;{\bf k}_{3};\omega_{3}}
\Psi^{\mathrm{cl}}_{\alpha;{\bf k}_{2};\omega_{2}}
\Psi^{\mathrm{cl}}_{\beta;{\bf k}_{4};\omega_{4}} 
\nonumber
\\
&
+
\bar{\Psi}^{\mathrm{cl}}_{\alpha;{\bf k}_{1};\omega_{1}} 
\bar{\Psi}^{\mathrm{cl}}_{\beta;{\bf k}_{3};\omega_{3}}
\Psi^{\mathrm{q}}_{\alpha;{\bf k}_{2};\omega_{2}}
\Psi^{\mathrm{cl}}_{\beta;{\bf k}_{4};\omega_{4}} 
+
\bar{\Psi}^{\mathrm{q}}_{\alpha;{\bf k}_{1};\omega_{1}} 
\bar{\Psi}^{\mathrm{cl}}_{\beta;{\bf k}_{3};\omega_{3}}
\Psi^{\mathrm{cl}}_{\alpha;{\bf k}_{2};\omega_{2}}
\Psi^{\mathrm{cl}}_{\beta;{\bf k}_{4};\omega_{4}} 
\rangle
\delta_{{\bf k}_{1}-{\bf k}_{2},{\bf k}_{4}-{\bf k}_{3}} 
\delta_{\omega_{1}-\omega_{2},\omega_{4}-\omega_{3}}
\\
=
&
i\frac{J}{2}
\int_{t}
\int_{\bf k}
I_{3}({\bf k})
\bar{\Psi}^{\mathrm{cl}}_{\alpha;{\bf k}}(t) 
\Psi^{\mathrm{q}}_{\beta;{\bf k}}(t) 
+
i\frac{J}{2} 
I_{1}
\int_{t}
\int_{\bf k}
\gamma_{0}
\bar{\Psi}^{\mathrm{cl}}_{\beta;{\bf k}}(t) 
\Psi^{\mathrm{q}}_{\beta;{\bf k}}(t) 
\\
+
&
i\frac{J}{2}
\int_{t}
\int_{\bf k}
I_{3}^{*}({\bf k})
\bar{\Psi}^{\mathrm{q}}_{\beta;{\bf k}}(t) 
\Psi^{\mathrm{cl}}_{\alpha;{\bf k}}(t) 
+
i\frac{J}{2}
I_{1}
\int_{t}
\int_{\bf k}
\gamma_{0}
\bar{\Psi}^{\mathrm{q}}_{\beta;{\bf k}}(t) 
\Psi^{\mathrm{cl}}_{\beta;{\bf k}}(t) 
\\
+
&
i\frac{J}{2}
\int_{\bf k}
I_{3}^{*}({\bf k})
\bar{\Psi}^{\mathrm{cl}}_{\beta;{\bf k}}(t) 
\Psi^{\mathrm{q}}_{\alpha;{\bf k}}(t) 
+
i\frac{J}{2}
I_{1}
\int_{t}
\int_{\bf k}
\gamma_{0}
\bar{\Psi}^{\mathrm{q}}_{\alpha;{\bf k}}(t) 
\Psi^{\mathrm{cl}}_{\alpha;{\bf k}}(t) 
\\
+
&
i\frac{J}{2}
\int_{t}
\int_{\bf k}
I_{3}({\bf k})
\bar{\Psi}^{\mathrm{q}}_{\alpha;{\bf k}}(t) 
\Psi^{\mathrm{cl}}_{\beta;{\bf k}}(t)
+
i\frac{J}{2}
I_{1}
\int_{t}
\int_{\bf k}
\gamma_{0}
\bar{\Psi}^{\mathrm{q}}_{\alpha;{\bf k}}(t) 
\Psi^{\mathrm{cl}}_{\alpha;{\bf k}}(t) 
\end{align}
New integral appeared above is
\begin{align}
I_{3}({\bf k}) = \int_{\bf q}\int_{\epsilon}iG_{\alpha\beta}^{\mathrm{K}}(\epsilon;{\bf q})\gamma_{{\bf k}-{\bf q}}
=
-\int_{\bf q}\frac{\gamma_{{\bf k}-{\bf q}} \gamma_{{\bf q}}}{ \vert \gamma_{{\bf q}} \vert}
\left[ n_{\mathrm{B}}(\epsilon_{+;{\bf q}}) - n_{\mathrm{B}}(\epsilon_{-;{\bf q}})\right].
\end{align}
Overall, we have for the Hartree-Fock corrections
\begin{align}
\sum_{j=1}^{5}\langle iS_{\mathrm{interaction};j}\rangle 
=
&
-i
\frac{J}{2}
\int_{t}
\int_{{\bf k}}
\left[I_{1}\gamma_{{\bf k}} - I_{3}({\bf k})\right]
\bar{\Psi}^{\mathrm{cl}}_{\alpha;{\bf k}}(t)
\Psi^{\mathrm{q}}_{\beta;{\bf k}}(t)
-i
\frac{J}{2}
(I_{2} - I_{1}\gamma_{0})
\int_{t}
\int_{\bf k}
\bar{\Psi}^{\mathrm{cl}}_{\alpha;{\bf k}}(t)
\Psi^{\mathrm{q}}_{\alpha;{\bf k}}(t)
\\
&
-i
\frac{J}{2}
(I_{2} - I_{1}\gamma_{0})
\int_{t}
\int_{\bf k}
\bar{\Psi}^{\mathrm{q}}_{\alpha;{\bf k}}(t)
\Psi^{\mathrm{cl}}_{\alpha;{\bf k}}(t)
-i
\frac{J}{2}
\int_{t}
\int_{{\bf k}}
\left[I_{1}\gamma_{{\bf k}} - I_{3}({\bf k})\right]
\bar{\Psi}^{\mathrm{q}}_{\alpha;{\bf k}}(t)
\Psi^{\mathrm{cl}}_{\beta;{\bf k}}(t)
\\
&
-i
\frac{J}{2}
\int_{t}
\int_{{\bf k}}
\left[I_{1}\gamma_{{\bf k}}^{*} - I_{3}^{*}({\bf k})\right]
\bar{\Psi}^{\mathrm{cl}}_{\beta;{\bf k}}(t)
\Psi^{\mathrm{q}}_{\alpha;{\bf k}}(t)
-i
\frac{J}{2}
(I_{2} - I_{1}\gamma_{0})
\int_{t}
\int_{\bf k}
\bar{\Psi}^{\mathrm{cl}}_{\beta;{\bf k}}(t)
\Psi^{\mathrm{q}}_{\beta;{\bf k}}(t)
\\
&
-
i
\frac{J}{2}
(I_{2} - I_{1}\gamma_{0})
\int_{t}
\int_{\bf k}
\bar{\Psi}^{\mathrm{q}}_{\beta;{\bf k}}(t)
\Psi^{\mathrm{cl}}_{\beta;{\bf k}}(t)
-i
\frac{J}{2}
\int_{t}
\int_{{\bf k}}
\left[I_{1}\gamma_{{\bf k}}^{*} - I_{3}^{*}({\bf k})\right]
\bar{\Psi}^{\mathrm{q}}_{\beta;{\bf k}}(t)
\Psi^{\mathrm{cl}}_{\alpha;{\bf k}}(t).
\end{align}
Integrals are
\begin{align}
I_{2} - I_{1}\gamma_{0} =
 -\int_{\bf q}\left( \gamma_{0} + \vert \gamma_{\bf q}\vert \right) n_{\mathrm{B}}(\epsilon_{+;{\bf q}})
 -\int_{\bf q}\left( \gamma_{0} - \vert \gamma_{\bf q}\vert \right) n_{\mathrm{B}}(\epsilon_{-;{\bf q}})
\approx -\left( \frac{T}{3SJ}\right)^2 \frac{\pi}{2}\gamma_{0},
\end{align}
and 
\begin{align}
I_{1}\gamma_{\bf k} - I_{3}({\bf k})
=\int_{\bf q}\left(\gamma_{\bf k} + \frac{\gamma_{{\bf k}-{\bf q}}\gamma_{\bf q}}{\vert \gamma_{\bf q}\vert} \right)n_{\mathrm{B}}(\epsilon_{+;{\bf q}})
+
\int_{\bf q}\left(\gamma_{\bf k} - \frac{\gamma_{{\bf k}-{\bf q}}\gamma_{\bf q}}{\vert \gamma_{\bf q}\vert} \right)n_{\mathrm{B}}(\epsilon_{-;{\bf q}})
\approx \left( \frac{T}{3SJ}\right)^2 \frac{\pi}{2}\gamma_{{\bf k}},
\end{align}
which are approximated at low temperatures, $T<3SJ$, under assumption that only the $\epsilon_{-;{\bf q}}$ magnon band contributes to the integrals. At temperatures $T\sim 3SJ$ (in the vicinity of the Curie temperature) both magnon bands will contribute, and, hence, the magnitude of integrals increase.

We then get Hartree-Fock corrected Hamiltonian describing the magnons 
\begin{align}
\hat{H} = 
JS\left[1-\frac{\pi}{4S}\left( \frac{T}{3SJ}\right)^2 \right]
\left[ 
\begin{array}{cc}
3 & - \gamma_{\bf k} \\
- \gamma_{\bf k}^{*} & 3
\end{array}
\right] 
\equiv
\tilde{J}S\left[ 
\begin{array}{cc}
3 & - \gamma_{\bf k} \\
- \gamma_{\bf k}^{*} & 3
\end{array}
\right] ,
\end{align}
where $\tilde{J} = J\left[1-\frac{\pi}{4S}\left( \frac{T}{3SJ}\right)^2 \right]$.
Exactly this Hamiltonian will be used below when calculating the ladder equation.

\subsection{Instability due to pumping}
We neglect the Hartree-Fock corrections by setting $T=0$.
Collecting all generated pumping terms, we construct a secular equation for $\Omega = 3SJ$,
\begin{align}\label{secularSM}
\mathrm{det}
\left[ 
\begin{array}{cccc}
\Omega+\epsilon - 3SJ & SJ\gamma_{\bf k} & -\Delta^2\gamma_{0} & \Delta^2\gamma_{{\bf k}} \\
SJ\gamma_{\bf k}^{*} & \Omega+\epsilon - 3SJ  & \Delta^2\gamma^{*}_{{\bf k}} & -\Delta^2\gamma_{0} \\
-\Delta^2\gamma_{0} & \Delta^2\gamma_{{\bf k}} & \Omega - \epsilon - 3SJ & SJ\gamma_{\bf k} \\
\Delta^2\gamma^{*}_{{\bf k}} & -\Delta^2\gamma_{0} & SJ\gamma^{*}_{\bf k} & \Omega - \epsilon - 3SJ
\end{array}
\right] 
= 0.
\end{align}
The Hamiltonian is similar to that of the BdG model, but only due to the presence of the anomalous terms.
The frequency structure is different because of the boson commutation relation the fields obey in our case.
We get
\begin{align}
\epsilon_{\pm}^{2} =  \left(\Omega - 3SJ \pm SJ\vert \gamma_{\bf k} \vert \right)^2 - \Delta^4(\gamma_{0} \mp \vert \gamma_{\bf k}\vert)^2.
\end{align}

Let us study the effect of Dzyaloshinskii-Moriya interaction Eq. (\ref{DMI_Hamiltonian}) on the magnon pairing in the vicinity of the Dirac points, i.e. for $\zeta = 0$. This is motivated by the fact that the DMI is the largest at the Dirac points.
The secular equation is now
\begin{align}\label{secularDMI}
\mathrm{det}
&
\left[ 
\begin{array}{cccc}
\chi+\epsilon  & SJ\gamma_{\bf k} & -3\Delta^2 & 0 \\
SJ\gamma_{\bf k}^{*} & -\chi+\epsilon   & 0 & -3\Delta^2 \\
-3\Delta^2 & 0 & \chi - \epsilon  & SJ\gamma_{\bf k} \\
0 & -3\Delta^2 & SJ\gamma_{\bf k}^{*} & -\chi - \epsilon 
\end{array}
\right] 
= 0,
\end{align}
where $\chi =3\sqrt{3}SD$, and $\vert \gamma_{\bf k} \vert \approx \frac{\sqrt{3}}{2}k$. The spectrum of magnon pairs is now 
\begin{align}
\epsilon^2_{\pm} = (SJ)^2 \frac{3}{4}k^2 + \chi^2  - 9\Delta^4.
\end{align}
We conclude that if $\vert \chi \vert  \geq  3\Delta^2$ there will be no instability in the system. In unpumped ferromagnet such Dzyaloshinskii-Moriya interaction opens up a gap at the Dirac points in the spectrum of the magnons. Then, for the Dirac magnons paired state to occur, pumping should overcome this gap.

\subsection{Ladder equation}

\begin{figure}[h] 
\centerline{
\includegraphics[width=0.35\textwidth,height=0.15\textheight]{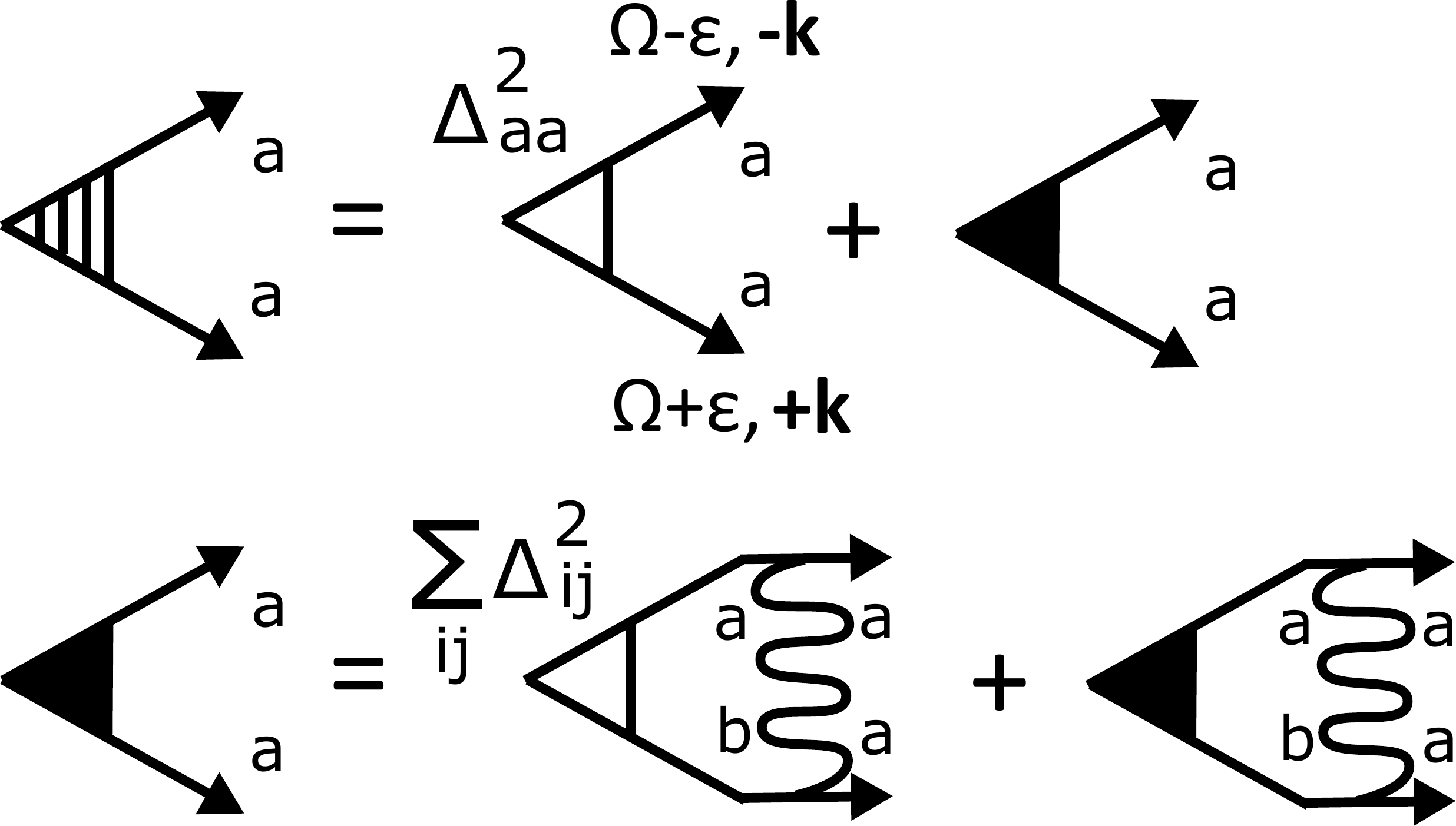}
}
\protect\caption{Graphic equation for the pairing interaction strength. Here empty triangle stands for the initial pairing interaction strength $\Delta^2_{ij}$ defined in accordance with Eq. (\ref{secularSM}), $\Delta^2_{\mathrm{aa}}=\Delta^2_{\mathrm{bb}} =- \Delta^2 \gamma_{0}$, and $\Delta^2_{\mathrm{ab}} =(\Delta^2_{\mathrm{ba}})^{*} = \Delta^2\gamma_{\bf k}$. 
Black tringle is intermediately renormalized pairing interaction strength, and the wavy lines stand for the interaction. 
Lined triangle is the overall renormalized pairing interaction strength.}

\label{fig:ladderSM}  

\end{figure}

The action describing the pump is
\begin{align}
iS_{\mathrm{pump}} = -i\int_{t} H_{\mathrm{pump}} 
\rightarrow
&
 - i
\int_{\{\epsilon\}}   \int_{\{{\bf p}\}}   
\frac{J}{4}
3
\left(\frac{\Gamma \sqrt{S}}{3SJ}\right)^2
\bar{\Psi}^{\mathrm{cl}}_{\alpha;{\bf p}_{1};\epsilon_{1}} 
\bar{\Psi}^{\mathrm{q}}_{\alpha;{\bf p}_{3};\epsilon_{3}}
\delta_{{\bf p}_{1},-{\bf p}_{3}}
\delta_{\epsilon_{1}-\Omega,\Omega-\epsilon_{3}}
\label{pump1}
\\
&
 + i
\int_{\{\epsilon\}}   \int_{\{{\bf p}\}}   
\frac{J}{4}
\gamma_{{\bf p}_{1}}
\left(\frac{\Gamma \sqrt{S}}{3SJ}\right)^2
\bar{\Psi}^{\mathrm{cl}}_{\alpha;{\bf p}_{1};\epsilon_{1}} 
\bar{\Psi}^{\mathrm{q}}_{\beta;{\bf p}_{3};\epsilon_{3}}
\delta_{{\bf p}_{1},-{\bf p}_{3}}
\delta_{\epsilon_{1}-\Omega,\Omega-\epsilon_{3}}
\label{pump2}
\\
&
 - i
\int_{\{\epsilon\}}   \int_{\{{\bf p}\}}   
\frac{J}{4}
3
\left(\frac{\Gamma \sqrt{S}}{3SJ}\right)^2
\bar{\Psi}^{\mathrm{cl}}_{\beta;{\bf p}_{1};\epsilon_{1}} 
\bar{\Psi}^{\mathrm{q}}_{\beta;{\bf p}_{3};\epsilon_{3}}
\delta_{{\bf p}_{1},-{\bf p}_{3}}
\delta_{\epsilon_{1}-\Omega,\Omega-\epsilon_{3}}
\label{pump3}
\\
&
 + i
\int_{\{\epsilon\}}   \int_{\{{\bf p}\}}   
\frac{J}{4}
\gamma_{{\bf p}_{1}}^{*}
\left(\frac{\Gamma \sqrt{S}}{3SJ}\right)^2
\bar{\Psi}^{\mathrm{cl}}_{\beta;{\bf p}_{1};\epsilon_{1}} 
\bar{\Psi}^{\mathrm{q}}_{\alpha;{\bf p}_{3};\epsilon_{3}}
\delta_{{\bf p}_{1},-{\bf p}_{3}}
\delta_{\epsilon_{1}-\Omega,\Omega-\epsilon_{3}}
\label{pump4}
\end{align}
where by right arrow we mean picking a particular term from the overall expression. 
Below, as an example, we wish to see how structure of Eq. (\ref{pump1}) gets renormalized by the interactions. 
For that we construct a ladder equation shown in Fig. \ref{fig:ladderSM}.
It turns out that only 
\begin{align}\label{interactionSM}
iS_{\mathrm{interaction}} = -i\int_{t} H_{\mathrm{interaction}} 
\rightarrow -i \frac{J}{4} \int_{\{\omega\}}  \int_{\{{\bf k}\}}  \gamma_{{\bf k}_{4}}
\bar{\Psi}^{\mathrm{cl}}_{\alpha;{\bf k}_{1};\omega_{1}} 
\bar{\Psi}^{\mathrm{q}}_{\alpha;{\bf k}_{3};\omega_{3}}
\Psi^{\mathrm{cl}}_{\alpha;{\bf k}_{2};\omega_{2}}
\Psi^{\mathrm{cl}}_{\beta;{\bf k}_{4};\omega_{4}} 
\delta_{{\bf k}_{1}-{\bf k}_{2},{\bf k}_{4}-{\bf k}_{3}} 
\delta_{\omega_{1}-\omega_{2},\omega_{4}-\omega_{3}}
\end{align}
part of the interaction can reproduce selected by us part of the pump. 
Contraction of the interaction Eq. (\ref{interactionSM}) with the first term, namely Eq. (\ref{pump1}), in the pump's Hamiltonian, gives the following expression
\begin{align}
&
\langle
\bar{\Psi}^{\mathrm{cl}}_{\alpha;{\bf k}_{1};\omega_{1}} 
\bar{\Psi}^{\mathrm{q}}_{\alpha;{\bf k}_{3};\omega_{3}}
\Psi^{\mathrm{cl}}_{\alpha;{\bf k}_{2};\omega_{2}}
\Psi^{\mathrm{cl}}_{\beta;{\bf k}_{4};\omega_{4}} 
\bar{\Psi}^{\mathrm{cl}}_{\alpha;{\bf p}_{1};\epsilon_{1}} 
\bar{\Psi}^{\mathrm{q}}_{\alpha;{\bf p}_{3};\epsilon_{3}}
\rangle
\\
&
=
\langle
\Psi^{\mathrm{cl}}_{\beta;{\bf k}_{4};\omega_{4}} 
\bar{\Psi}^{\mathrm{cl}}_{\alpha;{\bf p}_{1};\epsilon_{1}} 
\rangle
\langle
\Psi^{\mathrm{cl}}_{\alpha;{\bf k}_{2};\omega_{2}}
\bar{\Psi}^{\mathrm{q}}_{\alpha;{\bf p}_{3};\epsilon_{3}}
\rangle
\bar{\Psi}^{\mathrm{cl}}_{\alpha;{\bf k}_{1};\omega_{1}} 
\bar{\Psi}^{\mathrm{q}}_{\alpha;{\bf k}_{3};\omega_{3}}
+
\langle
\Psi^{\mathrm{cl}}_{\alpha;{\bf k}_{2};\omega_{2}}
\bar{\Psi}^{\mathrm{cl}}_{\alpha;{\bf p}_{1};\epsilon_{1}} 
\rangle
\langle
\Psi^{\mathrm{cl}}_{\beta;{\bf k}_{4};\omega_{4}} 
\bar{\Psi}^{\mathrm{q}}_{\alpha;{\bf p}_{3};\epsilon_{3}}
\rangle
\bar{\Psi}^{\mathrm{cl}}_{\alpha;{\bf k}_{1};\omega_{1}} 
\bar{\Psi}^{\mathrm{q}}_{\alpha;{\bf k}_{3};\omega_{3}}
\\
&
=
-
[ 
G^{\mathrm{K}}_{\beta\alpha}({\bf k}_{4};\omega_{4})
G^{\mathrm{R}}_{\alpha\alpha}({\bf k}_{2};\omega_{2})\delta_{{\bf k}_{4},{\bf p}_{1}}\delta_{\omega_{4},\epsilon_{1}}
\delta_{{\bf k}_{2},{\bf p}_{3}}\delta_{\omega_{2},\epsilon_{3}}
+ 
G^{\mathrm{R}}_{\beta\alpha}({\bf k}_{4};\omega_{2})
G^{\mathrm{K}}_{\alpha\alpha}({\bf k}_{2};\omega_{4})
\delta_{{\bf k}_{2},{\bf p}_{1}}\delta_{\omega_{2},\epsilon_{1}}
\delta_{{\bf k}_{4},{\bf p}_{3}}\delta_{\omega_{4},\epsilon_{3}}
]
\bar{\Psi}^{\mathrm{cl}}_{\alpha;{\bf k}_{1};\omega_{1}} 
\bar{\Psi}^{\mathrm{q}}_{\alpha;{\bf k}_{3};\omega_{3}}.
\end{align}
Contraction of the interaction Eq. (\ref{interactionSM}) with the second term in the pump's Hamiltonian, namely Eq. (\ref{pump2}), results in the following expression
\begin{align}
&
\gamma_{{\bf p}_{1}}
\langle
\bar{\Psi}^{\mathrm{cl}}_{\alpha;{\bf k}_{1};\omega_{1}} 
\bar{\Psi}^{\mathrm{q}}_{\alpha;{\bf k}_{3};\omega_{3}}
\Psi^{\mathrm{cl}}_{\alpha;{\bf k}_{2};\omega_{2}}
\Psi^{\mathrm{cl}}_{\beta;{\bf k}_{4};\omega_{4}} 
\bar{\Psi}^{\mathrm{cl}}_{\alpha;{\bf p}_{1};\epsilon_{1}} 
\bar{\Psi}^{\mathrm{q}}_{\beta;{\bf p}_{3};\epsilon_{3}}
\rangle
\\
&
=
-
\gamma_{{\bf p}_{1}}
[ 
G^{\mathrm{K}}_{\alpha\alpha}({\bf k}_{2};\omega_{2})
G^{\mathrm{R}}_{\beta\beta}({\bf k}_{4};\omega_{4})
\delta_{{\bf k}_{2},{\bf p}_{1}}\delta_{\omega_{2},\epsilon_{1}}
\delta_{{\bf k}_{4},{\bf p}_{3}}\delta_{\omega_{4},\epsilon_{3}}
+ 
G^{\mathrm{R}}_{\alpha\beta}({\bf k}_{2};\omega_{2})
G^{\mathrm{K}}_{\beta\alpha}({\bf k}_{4};\omega_{4})
\delta_{{\bf k}_{2},{\bf p}_{3}}\delta_{\omega_{2},\epsilon_{3}}
\delta_{{\bf k}_{4},{\bf p}_{1}}\delta_{\omega_{4},\epsilon_{1}}
]
\bar{\Psi}^{\mathrm{cl}}_{\alpha;{\bf k}_{1};\omega_{1}} 
\bar{\Psi}^{\mathrm{q}}_{\alpha;{\bf k}_{3};\omega_{3}}.
\end{align}
Contraction of the interaction Eq. (\ref{interactionSM}) with the third term, namely Eq. (\ref{pump3}), in the pump's Hamiltonian, gives the following bracket
\begin{align}
&
\langle
\bar{\Psi}^{\mathrm{cl}}_{\alpha;{\bf k}_{1};\omega_{1}} 
\bar{\Psi}^{\mathrm{q}}_{\alpha;{\bf k}_{3};\omega_{3}}
\Psi^{\mathrm{cl}}_{\alpha;{\bf k}_{2};\omega_{2}}
\Psi^{\mathrm{cl}}_{\beta;{\bf k}_{4};\omega_{4}} 
\bar{\Psi}^{\mathrm{cl}}_{\beta;{\bf p}_{1};\epsilon_{1}} 
\bar{\Psi}^{\mathrm{q}}_{\beta;{\bf p}_{3};\epsilon_{3}}
\rangle
\\
&
=
\langle
\Psi^{\mathrm{cl}}_{\beta;{\bf k}_{4};\omega_{4}} 
\bar{\Psi}^{\mathrm{cl}}_{\beta;{\bf p}_{1};\epsilon_{1}} 
\rangle
\langle
\Psi^{\mathrm{cl}}_{\alpha;{\bf k}_{2};\omega_{2}}
\bar{\Psi}^{\mathrm{q}}_{\beta;{\bf p}_{3};\epsilon_{3}}
\rangle
\bar{\Psi}^{\mathrm{cl}}_{\alpha;{\bf k}_{1};\omega_{1}} 
\bar{\Psi}^{\mathrm{q}}_{\alpha;{\bf k}_{3};\omega_{3}}
+
\langle
\Psi^{\mathrm{cl}}_{\alpha;{\bf k}_{2};\omega_{2}}
\bar{\Psi}^{\mathrm{cl}}_{\beta;{\bf p}_{1};\epsilon_{1}} 
\rangle
\langle
\Psi^{\mathrm{cl}}_{\beta;{\bf k}_{4};\omega_{4}} 
\bar{\Psi}^{\mathrm{q}}_{\beta;{\bf p}_{3};\epsilon_{3}}
\rangle
\bar{\Psi}^{\mathrm{cl}}_{\alpha;{\bf k}_{1};\omega_{1}} 
\bar{\Psi}^{\mathrm{q}}_{\alpha;{\bf k}_{3};\omega_{3}}
\\
&
=
-
[ 
G^{\mathrm{K}}_{\beta\beta}({\bf k}_{4};\omega_{4})
G^{\mathrm{R}}_{\alpha\beta}({\bf k}_{2};\omega_{2})\delta_{{\bf k}_{4},{\bf p}_{1}}\delta_{\omega_{4},\epsilon_{1}}
\delta_{{\bf k}_{2},{\bf p}_{3}}\delta_{\omega_{2},\epsilon_{3}}
+ 
G^{\mathrm{R}}_{\beta\beta}({\bf k}_{4};\omega_{2})
G^{\mathrm{K}}_{\alpha\beta}({\bf k}_{2};\omega_{4})
\delta_{{\bf k}_{2},{\bf p}_{1}}\delta_{\omega_{2},\epsilon_{1}}
\delta_{{\bf k}_{4},{\bf p}_{3}}\delta_{\omega_{4},\epsilon_{3}}
]
\bar{\Psi}^{\mathrm{cl}}_{\alpha;{\bf k}_{1};\omega_{1}} 
\bar{\Psi}^{\mathrm{q}}_{\alpha;{\bf k}_{3};\omega_{3}}.
\end{align}
Contraction of the interaction Eq. (\ref{interactionSM}) with the second term in the pump's Hamiltonian, namely Eq. (\ref{pump4}), results in the following expression
\begin{align}
&
\gamma_{{\bf p}_{1}}^{*}
\langle
\bar{\Psi}^{\mathrm{cl}}_{\alpha;{\bf k}_{1};\omega_{1}} 
\bar{\Psi}^{\mathrm{q}}_{\alpha;{\bf k}_{3};\omega_{3}}
\Psi^{\mathrm{cl}}_{\alpha;{\bf k}_{2};\omega_{2}}
\Psi^{\mathrm{cl}}_{\beta;{\bf k}_{4};\omega_{4}} 
\bar{\Psi}^{\mathrm{cl}}_{\beta;{\bf p}_{1};\epsilon_{1}} 
\bar{\Psi}^{\mathrm{q}}_{\alpha;{\bf p}_{3};\epsilon_{3}}
\rangle
\\
&
=
-
\gamma_{{\bf p}_{1}}^{*}
[ 
G^{\mathrm{K}}_{\alpha\beta}({\bf k}_{2};\omega_{4})
G^{\mathrm{R}}_{\beta\alpha}({\bf k}_{4};\omega_{2})
\delta_{{\bf k}_{4},{\bf p}_{3}}\delta_{\omega_{4},\epsilon_{3}}
\delta_{{\bf k}_{2},{\bf p}_{1}}\delta_{\omega_{2},\epsilon_{1}}
+ 
G^{\mathrm{R}}_{\alpha\alpha}({\bf k}_{2};\omega_{2})
G^{\mathrm{K}}_{\beta\beta}({\bf k}_{4};\omega_{4})
\delta_{{\bf k}_{2},{\bf p}_{3}}\delta_{\omega_{2},\epsilon_{3}}
\delta_{{\bf k}_{4},{\bf p}_{1}}\delta_{\omega_{4},\epsilon_{1}}
]
\bar{\Psi}^{\mathrm{cl}}_{\alpha;{\bf k}_{1};\omega_{1}} 
\bar{\Psi}^{\mathrm{q}}_{\alpha;{\bf k}_{3};\omega_{3}}.
\end{align}
Summing all four contributions, we get
\begin{align}
&
\langle(iS_{\mathrm{interaction}})(iS_{\mathrm{pump}} )\rangle
\rightarrow
\left( \frac{J}{4}\right)^2
\left(\frac{\Gamma \sqrt{S}}{3SJ}\right)^2
\int_{\{\omega\}}  \int_{\{{\bf k}\}} 
\gamma_{{\bf k}_{4}}
\int_{\{\epsilon\}}   \int_{\{{\bf p}\}} 
\delta_{{\bf k}_{1}-{\bf k}_{2},{\bf k}_{4}-{\bf k}_{3}} 
\delta_{\omega_{1}-\omega_{2},\omega_{4}-\omega_{3}}
\delta_{{\bf p}_{1},-{\bf p}_{3}}
\delta_{\epsilon_{1}-\Omega,\Omega-\epsilon_{3}}
\\
&
\times
(
-3\langle
\bar{\Psi}^{\mathrm{cl}}_{\alpha;{\bf k}_{1};\omega_{1}} 
\bar{\Psi}^{\mathrm{q}}_{\alpha;{\bf k}_{3};\omega_{3}}
\Psi^{\mathrm{cl}}_{\alpha;{\bf k}_{2};\omega_{2}}
\Psi^{\mathrm{cl}}_{\beta;{\bf k}_{4};\omega_{4}} 
\bar{\Psi}^{\mathrm{cl}}_{\alpha;{\bf p}_{1};\epsilon_{1}} 
\bar{\Psi}^{\mathrm{q}}_{\alpha;{\bf p}_{3};\epsilon_{3}}
\rangle
+
\gamma_{{\bf p}_{1}}
\langle
\bar{\Psi}^{\mathrm{cl}}_{\alpha;{\bf k}_{1};\omega_{1}} 
\bar{\Psi}^{\mathrm{q}}_{\alpha;{\bf k}_{3};\omega_{3}}
\Psi^{\mathrm{cl}}_{\alpha;{\bf k}_{2};\omega_{2}}
\Psi^{\mathrm{cl}}_{\beta;{\bf k}_{4};\omega_{4}} 
\bar{\Psi}^{\mathrm{cl}}_{\alpha;{\bf p}_{1};\epsilon_{1}} 
\bar{\Psi}^{\mathrm{q}}_{\beta;{\bf p}_{3};\epsilon_{3}}
\rangle
\\
&
-3
\langle
\bar{\Psi}^{\mathrm{cl}}_{\alpha;{\bf k}_{1};\omega_{1}} 
\bar{\Psi}^{\mathrm{q}}_{\alpha;{\bf k}_{3};\omega_{3}}
\Psi^{\mathrm{cl}}_{\alpha;{\bf k}_{2};\omega_{2}}
\Psi^{\mathrm{cl}}_{\beta;{\bf k}_{4};\omega_{4}} 
\bar{\Psi}^{\mathrm{cl}}_{\beta;{\bf p}_{1};\epsilon_{1}} 
\bar{\Psi}^{\mathrm{q}}_{\beta;{\bf p}_{3};\epsilon_{3}}
\rangle
+
\gamma_{{\bf p}_{1}}^{*}
\langle
\bar{\Psi}^{\mathrm{cl}}_{\alpha;{\bf k}_{1};\omega_{1}} 
\bar{\Psi}^{\mathrm{q}}_{\alpha;{\bf k}_{3};\omega_{3}}
\Psi^{\mathrm{cl}}_{\alpha;{\bf k}_{2};\omega_{2}}
\Psi^{\mathrm{cl}}_{\beta;{\bf k}_{4};\omega_{4}} 
\bar{\Psi}^{\mathrm{cl}}_{\beta;{\bf p}_{1};\epsilon_{1}} 
\bar{\Psi}^{\mathrm{q}}_{\alpha;{\bf p}_{3};\epsilon_{3}}
\rangle
)
\\
&
=
\left( \frac{J}{4}\right)^2
\left(\frac{\Gamma \sqrt{S}}{3SJ}\right)^2
\bigg\{
\int_{{\bf k}}\int_{\epsilon}
3\gamma_{-{\bf k}}
\left[ G^{\mathrm{K}}_{\beta\alpha}(-{\bf k};\Omega- \epsilon) G^{\mathrm{R}}_{\alpha\alpha}({\bf k};\Omega+ \epsilon)
+
 G^{\mathrm{R}}_{\beta\alpha}(-{\bf k};\Omega- \epsilon) G^{\mathrm{K}}_{\alpha\alpha}({\bf k};\Omega+ \epsilon)
\right]
\\
&
~~~~~~~~~~~~~~~~~~~~~~~~~~~~~~
-
\int_{{\bf k}}\int_{\epsilon}
\left[ 
\vert \gamma_{\bf k}\vert^2
G^{\mathrm{K}}_{\alpha\alpha}({\bf k};\Omega + \epsilon) G^{\mathrm{R}}_{\beta\beta}(-{\bf k};\Omega - \epsilon)
+
\gamma_{\bf k}^2
 G^{\mathrm{K}}_{\beta\alpha}({\bf k};\Omega+ \epsilon) G^{\mathrm{R}}_{\alpha\beta}(-{\bf k};\Omega - \epsilon)
\right]
\bigg\}
\\
&
~~~~~~~~~
\times
\int_{\{ {\bf k}\}}\int_{\{ \omega \}}
\bar{\Psi}^{\mathrm{cl}}_{\alpha;{\bf k}_{1};\omega_{1}} 
\bar{\Psi}^{\mathrm{q}}_{\alpha;{\bf k}_{3};\omega_{3}}
\delta_{{\bf k}_{1},-{\bf k}_{3}}
\delta_{\omega_{1}-\Omega,\Omega-\omega_{3}}
\\
&
+
\left( \frac{J}{4}\right)^2
\left(\frac{\Gamma \sqrt{S}}{3SJ}\right)^2
\bigg\{
\int_{{\bf k}}\int_{\epsilon}
3\gamma_{-{\bf k}}
\left[ G^{\mathrm{K}}_{\beta\beta}(-{\bf k};\Omega- \epsilon) G^{\mathrm{R}}_{\alpha\beta}({\bf k};\Omega+ \epsilon)
+
 G^{\mathrm{R}}_{\beta\beta}(-{\bf k};\Omega- \epsilon) G^{\mathrm{K}}_{\alpha\beta}({\bf k};\Omega+ \epsilon)
\right]
\\
&
~~~~~~~~~~~~~~~~~~~~~~~~~~~~~~
-
\int_{{\bf k}}\int_{\epsilon}
\left[ 
\vert \gamma_{\bf k}\vert^2
G^{\mathrm{K}}_{\beta\beta}({\bf k};\Omega + \epsilon) G^{\mathrm{R}}_{\alpha\alpha}(-{\bf k};\Omega - \epsilon)
+
\gamma_{-\bf k}^2
 G^{\mathrm{K}}_{\alpha\beta}({\bf k};\Omega+ \epsilon) G^{\mathrm{R}}_{\beta\alpha}(-{\bf k};\Omega - \epsilon)
\right]
\bigg\}
\\
&
~~~~~~~~~
\times
\int_{\{ {\bf k}\}}\int_{\{ \omega \}}
\bar{\Psi}^{\mathrm{cl}}_{\alpha;{\bf k}_{1};\omega_{1}} 
\bar{\Psi}^{\mathrm{q}}_{\alpha;{\bf k}_{3};\omega_{3}}
\delta_{{\bf k}_{1},-{\bf k}_{3}}
\delta_{\omega_{1}-\Omega,\Omega-\omega_{3}}.
\end{align}
It can be shown that the two terms simply double each other. We will be using 
\begin{align}
G^{\mathrm{K}}({\bf k};\epsilon) = G^{\mathrm{R}}({\bf k} ;\epsilon){\cal F}_{\epsilon}- {\cal F}_{\epsilon}G^{\mathrm{A}}({\bf k}; \epsilon )
\end{align}
identity, 
and a generalization of $G^{\mathrm{R}}({\bf k};\epsilon)- G^{\mathrm{A}}({\bf k};\epsilon) = - 2\pi i \delta(\epsilon-\epsilon_{\bf k})$ identity for the honeycomb lattice.

\subsubsection{Case of $\Omega = 3S\tilde{J}$}
Let us calculate the step of the ladder for the $\Omega = 3S\tilde{J}$. Recall, that $\tilde{J}=J\left[1-\frac{\pi}{4S}\left( \frac{T}{3SJ}\right)^2 \right]$. First integral reads
\begin{align}
&
\int_{{\bf k}}\int_{\epsilon}
3\gamma_{-{\bf k}}
\left[ G^{\mathrm{K}}_{\beta\alpha}(-{\bf k};\Omega- \epsilon) G^{\mathrm{R}}_{\alpha\alpha}({\bf k};\Omega+ \epsilon)
+
 G^{\mathrm{R}}_{\beta\alpha}(-{\bf k};\Omega- \epsilon) G^{\mathrm{K}}_{\alpha\alpha}({\bf k};\Omega+ \epsilon)
\right]
\\
&
=
\frac{i}{2}
\int_{{\bf k}}
3\vert\gamma_{{\bf k}}\vert
\left[ 
\frac{{\cal F}_{\epsilon_{+;{\bf k}}} }{2\Omega - 6S\tilde{J} - 2S\tilde{J}\vert \gamma_{\bf k} \vert + i0} 
-
\frac{{\cal F}_{\epsilon_{-;{\bf k}}} }{2\Omega - 6S\tilde{J} + 2S\tilde{J}\vert \gamma_{\bf k} \vert + i0}   
\right]
= 
-\frac{i}{4S\tilde{J}}
\int_{\bf k} 3\left[ {\cal F}_{\epsilon_{+;{\bf k}}}  + {\cal F}_{\epsilon_{-;{\bf k}}} \right].
\end{align}
Here and below $\epsilon_{\pm {\bf k}} =\tilde{J}S\left( 3 \pm  \vert \gamma_{\bf k}\vert \right)$, unperturbed energy of the magnons. 
Second integrals reads
\begin{align}
&
\int_{{\bf k}}
\int_{\epsilon}
\left[ 
\vert \gamma_{\bf k}\vert^2
G^{\mathrm{K}}_{\alpha\alpha}({\bf k};\Omega + \epsilon) G^{\mathrm{R}}_{\beta\beta}(-{\bf k};\Omega - \epsilon)
+
\gamma_{\bf k}^2
 G^{\mathrm{K}}_{\beta\alpha}({\bf k};\Omega+ \epsilon) G^{\mathrm{R}}_{\alpha\beta}(-{\bf k};\Omega - \epsilon)
\right]
\\
&
= 
-
\frac{i}{2}
\int_{{\bf k}}
\vert\gamma_{{\bf k}}\vert^2
\left[ 
\frac{{\cal F}_{\epsilon_{+;{\bf k}}} }{2\Omega - 6S\tilde{J} - 2S\tilde{J}\vert \gamma_{\bf k} \vert + i0} 
+
\frac{{\cal F}_{\epsilon_{-;{\bf k}}} }{2\Omega - 6S\tilde{J} + 2S\tilde{J}\vert \gamma_{\bf k} \vert + i0}   
\right]
= 
\frac{i}{4S\tilde{J}}
\int_{\bf k}
\vert\gamma_{{\bf k}}\vert
\left[ {\cal F}_{\epsilon_{+;{\bf k}}}  - {\cal F}_{\epsilon_{-;{\bf k}}} \right].
\end{align}
Summing the two, we get
\begin{align}
&
\int_{{\bf k}}\int_{\epsilon}
3\gamma_{-{\bf k}}
\left[ G^{\mathrm{K}}_{\beta\alpha}(-{\bf k};\Omega- \epsilon) G^{\mathrm{R}}_{\alpha\alpha}({\bf k};\Omega+ \epsilon)
+
 G^{\mathrm{R}}_{\beta\alpha}(-{\bf k};\Omega- \epsilon) G^{\mathrm{K}}_{\alpha\alpha}({\bf k};\Omega+ \epsilon)
\right]
\\
-
&
\int_{{\bf k}}\int_{\epsilon}
\left[ 
\vert \gamma_{\bf k}\vert^2
G^{\mathrm{K}}_{\alpha\alpha}({\bf k};\Omega + \epsilon) G^{\mathrm{R}}_{\beta\beta}(-{\bf k};\Omega - \epsilon)
+
\gamma_{\bf k}^2
 G^{\mathrm{K}}_{\beta\alpha}({\bf k};\Omega+ \epsilon) G^{\mathrm{R}}_{\alpha\beta}(-{\bf k};\Omega - \epsilon)
\right]
\\
&
= - \frac{i}{4S\tilde{J}} \int_{\bf k}\left[ \left( 3+\vert \gamma_{\bf k} \vert \right){\cal F}_{\epsilon_{+;{\bf k}}} + \left( 3-\vert \gamma_{\bf k} \vert \right){\cal F}_{\epsilon_{-;{\bf k}}}\right]
\\
&
\approx
-\frac{i}{4S\tilde{J}} \left[ 6 +\frac{\pi}{3} \left(\frac{T}{S\tilde{J}}\right)^2 \right],
\end{align}
where 
\begin{align}
\int_{\bf k} \left[ \left( 3+\vert \gamma_{\bf k} \vert \right){\cal F}_{\epsilon_{+;{\bf k}}} + \left( 3-\vert \gamma_{\bf k} \vert \right){\cal F}_{\epsilon_{-;{\bf k}}}\right]
&
\approx
\int_{\bf k} \left[ 6 + 2 \frac{3-\vert \gamma_{\bf k} \vert}{e^{\frac{S\tilde{J}(3-\vert \gamma_{\bf k} \vert)}{T}} - 1}  \right]
=
24\sqrt{3} + \frac{1}{8\pi}\left(\frac{4T}{S\tilde{J}}\right)^2 
\int_{0}^{\infty} \frac{zdz}{e^{z}-1} 
\\
&
= 6 + 3\pi\left(\frac{T}{3S\tilde{J}}\right)^2 ,
\end{align}
where $\int_{\bf k} 6 =\frac{6}{(2\pi)^2} \int_{0}^{2\pi}dk_{x}\int_{0}^{2\pi}dk_{y}= 6$ is an integral over the period of the magnon's dispersion defined by $\gamma_{\bf k} = 2e^{i\frac{k_{x}}{2\sqrt{3}}}\cos\left( \frac{k_{y}}{2}\right)+ e^{-i\frac{k_{x}}{\sqrt{3}}}$. The integral counts all available for pairing magnon states.
Second term above can be neglected as it is always small, $T \ll SJ$. 
We used 
\begin{align}
{\cal F}_{\epsilon} = \coth\left(\frac{\epsilon}{2T}\right) = 1 + \frac{2}{e^{\frac{\epsilon}{T}}-1},
\end{align}
and 
\begin{align}
\left( 3+\vert \gamma_{\bf k} \vert \right){\cal F}_{\epsilon_{+;{\bf k}}} + \left( 3-\vert \gamma_{\bf k} \vert \right){\cal F}_{\epsilon_{-;{\bf k}}} 
= 6 + \frac{2\left( 3+\vert \gamma_{\bf k} \vert \right)}{e^{\frac{S\tilde{J} }{T}\left( 3+\vert \gamma_{\bf k} \vert \right)}-1} +  \frac{2\left( 3-\vert \gamma_{\bf k} \vert \right)}{e^{\frac{S\tilde{J}}{T}\left( 3-\vert \gamma_{\bf k} \vert \right)}-1} 
\approx 6+\frac{2\left( 3-\vert \gamma_{\bf k} \vert \right)}{e^{\frac{S\tilde{J}}{T}\left( 3-\vert \gamma_{\bf k} \vert \right)}-1} ,
\end{align}
which is a natural approximation, as only the low-energy magnons with $\epsilon_{-;{\bf k}}$ dispersion can contribute to the integral. The $\epsilon_{+;{\bf k}}$ are exponentially suppressed at small temperatures.
Then we have for the step of the ladder,
\begin{align}
\langle(iS_{\mathrm{interaction}})(iS_{\mathrm{pump}} )\rangle
&
\approx 
-2 \frac{i}{4S\tilde{J}}
\left( \frac{J}{4}\right)^2
\left(\frac{\Gamma \sqrt{S}}{3SJ}\right)^2
\left[ 6 + 3\pi\left(\frac{T}{3S\tilde{J}}\right)^2 \right] 
\int_{\{ {\bf k}\}}\int_{\{ \omega \}}
\bar{\Psi}^{\mathrm{cl}}_{\alpha;{\bf k}_{1};\omega_{1}} 
\bar{\Psi}^{\mathrm{q}}_{\alpha;{\bf k}_{3};\omega_{3}}
\delta_{{\bf k}_{1},-{\bf k}_{3}}
\delta_{\omega_{1}-\Omega,\Omega-\omega_{3}}
\end{align}
Summing the original pumping term, the first step of the ladder, and iterating the steps further, we get,
\begin{align}
&
iS_{\mathrm{pump}}  + \langle(iS_{\mathrm{interaction}})(iS_{\mathrm{pump}} )\rangle
\\
=
&
 -i3\frac{J}{4}
\left(\frac{\Gamma \sqrt{S}}{3SJ}\right)^2
\left[ 1+ \frac{1}{4S} \frac{J}{\tilde{J}} + \frac{\pi }{8S }\frac{J}{\tilde{J}} \left(\frac{T}{3S\tilde{J}}\right)^2 \right]
\int_{\{ {\bf k}\}}\int_{\{ \omega \}}
\bar{\Psi}^{\mathrm{cl}}_{\alpha;{\bf k}_{1};\omega_{1}} 
\bar{\Psi}^{\mathrm{q}}_{\alpha;{\bf k}_{3};\omega_{3}}
\delta_{{\bf k}_{1},-{\bf k}_{3}}
\delta_{\omega_{1}-\Omega,\Omega-\omega_{3}}
\\
&
\rightarrow
 -i3\frac{J}{4}
\left(\frac{\Gamma \sqrt{S}}{3SJ}\right)^2
\frac{1}{ 1- \frac{1}{4S} \frac{J}{\tilde{J}} - \frac{\pi }{8S }\frac{J}{\tilde{J}}\left(\frac{T}{3S\tilde{J}}\right)^2 }
\int_{\{ {\bf k}\}}\int_{\{ \omega \}}
\bar{\Psi}^{\mathrm{cl}}_{\alpha;{\bf k}_{1};\omega_{1}} 
\bar{\Psi}^{\mathrm{q}}_{\alpha;{\bf k}_{3};\omega_{3}}
\delta_{{\bf k}_{1},-{\bf k}_{3}}
\delta_{\omega_{1}-\Omega,\Omega-\omega_{3}},
\end{align}
clearly there is an enhancement of pairing.

\subsubsection{Case of $\Omega \neq 3SJ$}
Here we demonstrate that 
for $\Omega \neq 3SJ$ each step of the ladder acquires an imaginary part. 
Besides, we are going to show that the pumping gets suppressed by the rescattering processes described by the ladder as the frequency approaches $6SJ$. 
To see the general tendency of the renormalization of the pairing strength away from the Dirac points, we disregard Hartree-Fock corrections to the magnon dispersion.
We have for the step of the ladder,
\begin{align}
&
M(\Omega) \equiv \int_{{\bf k}}\int_{\epsilon}
3\gamma_{-{\bf k}}
\left[ G^{\mathrm{K}}_{\beta\alpha}(-{\bf k};\Omega- \epsilon) G^{\mathrm{R}}_{\alpha\alpha}({\bf k};\Omega+ \epsilon)
+
 G^{\mathrm{R}}_{\beta\alpha}(-{\bf k};\Omega- \epsilon) G^{\mathrm{K}}_{\alpha\alpha}({\bf k};\Omega+ \epsilon)
\right]
\\
&
-
\int_{{\bf k}}\int_{\epsilon}
\left[ 
\vert \gamma_{\bf k}\vert^2
G^{\mathrm{K}}_{\alpha\alpha}({\bf k};\Omega + \epsilon) G^{\mathrm{R}}_{\beta\beta}(-{\bf k};\Omega - \epsilon)
+
\gamma_{\bf k}^2
 G^{\mathrm{K}}_{\beta\alpha}({\bf k};\Omega+ \epsilon) G^{\mathrm{R}}_{\alpha\beta}(-{\bf k};\Omega - \epsilon)
\right]
\\
&
=
\frac{i}{2}
\int_{{\bf k}}
\vert\gamma_{{\bf k}}\vert
\left[ 
\frac{(3+\vert\gamma_{{\bf k}}\vert){\cal F}_{\epsilon_{+;{\bf k}}} }{2\Omega - 6SJ - 2SJ\vert \gamma_{\bf k} \vert + i0} 
-
\frac{(3-\vert\gamma_{{\bf k}}\vert){\cal F}_{\epsilon_{-;{\bf k}}} }{2\Omega - 6SJ + 2SJ\vert \gamma_{\bf k} \vert + i0}   
\right]
\\
&
=
\frac{i}{4}
\mathrm{PV}
\int_{{\bf k}}
\vert\gamma_{{\bf k}}\vert
\left[ 
\frac{(3+\vert\gamma_{{\bf k}}\vert){\cal F}_{\epsilon_{+;{\bf k}}} }{\zeta- SJ\vert \gamma_{\bf k} \vert } 
-
\frac{(3-\vert\gamma_{{\bf k}}\vert){\cal F}_{\epsilon_{-;{\bf k}}} }{\zeta + SJ\vert \gamma_{\bf k} \vert }   
\right]
\\
&
+\frac{i}{4}\left(-\frac{i\pi}{2}\right) \int_{{\bf k}} \vert\gamma_{{\bf k}}\vert \delta( \zeta- SJ\vert \gamma_{\bf k} \vert  )
(3+\vert\gamma_{{\bf k}}\vert){\cal F}_{\epsilon_{+;{\bf k}}} 
-\frac{i}{4}\left(-\frac{i\pi}{2}\right) \int_{{\bf k}} \vert\gamma_{{\bf k}}\vert \delta( \zeta +SJ\vert \gamma_{\bf k} \vert  )
(3-\vert\gamma_{{\bf k}}\vert){\cal F}_{\epsilon_{-;{\bf k}}} ,
\end{align}
where $\mathrm{PV}$ is the principal value of the integral, and where  $\zeta=\Omega-3SJ$. The imaginary part for $\zeta >0$ is evaluated as
\begin{align}
&
-\frac{i\pi}{2} \int_{{\bf k}} \vert\gamma_{{\bf k}}\vert \delta( \zeta- SJ\vert \gamma_{\bf k} \vert  )
(3+\vert\gamma_{{\bf k}}\vert){\cal F}_{\epsilon_{+;{\bf k}}} 
+\frac{i\pi}{2} \int_{{\bf k}} \vert\gamma_{{\bf k}}\vert \delta( \zeta +SJ\vert \gamma_{\bf k} \vert  )
(3-\vert\gamma_{{\bf k}}\vert){\cal F}_{\epsilon_{-;{\bf k}}} 
\\
&
= -\frac{i\pi}{2}\frac{\zeta}{(SJ)^2}\left(3+\frac{\zeta}{SJ}\right)
{\cal F}\left(3SJ+\zeta\right)\int_{\bf k}\delta( \zeta -SJ\vert \gamma_{\bf k} \vert  ),
\end{align}
where we kept the integral as it is.
The imaginary part is non-zero and works towards weakening of the pairing between magnons.

Let us estimate the step of the ladder when the pump frequency is $\Omega = 6SJ - \alpha$ and $\alpha$ is small.
Then $\zeta = 3SJ  - \alpha$, we approximate $\vert \gamma_{\bf k}\vert \approx 3 -\frac{k^2}{4}$, and we write for the step of the ladder
\begin{align}
&
\mathrm{Im}\left[M(\Omega)\right]
 \approx
\frac{1}{4}
\mathrm{PV}
\int_{{\bf k}}
\vert\gamma_{{\bf k}}\vert
\frac{(3+\vert\gamma_{{\bf k}}\vert){\cal F}_{\epsilon_{+;{\bf k}}} }{\zeta- SJ\vert \gamma_{\bf k} \vert } 
\approx
\frac{9}{2}
\mathrm{PV}
\int_{{\bf k}}
\frac{1}{-\alpha + SJ\frac{k^2}{4} } 
= \frac{9}{2\pi SJ} \mathrm{PV}\int_{0}^{\Lambda^2} \frac{dz}{z -\alpha \frac{4}{SJ}} 
\approx  \frac{9}{2\pi SJ} \ln\left(\frac{SJ \Lambda^2 }{4\alpha} \right),
\\
&
\mathrm{Re}\left[ M(\Omega)\right]
\approx
\frac{\pi}{4} \int_{{\bf k}} \vert\gamma_{{\bf k}}\vert \delta( \zeta- SJ\vert \gamma_{\bf k} \vert  )
(3+\vert\gamma_{{\bf k}}\vert){\cal F}_{\epsilon_{+;{\bf k}}} 
\approx 
\frac{9}{2SJ}.
\end{align}
We then get for the renormalization of the pairing
\begin{align}
\Delta^2 \rightarrow \frac{\Delta^2}{1+ \frac{3}{16\pi S} \ln\left(\frac{SJ \Lambda^2 }{4\alpha} \right) + i\frac{3}{16S}}.
\end{align}
Importantly, for frequencies away from the Dirac points, the structure of the renormalization due to the rescattering processes drastically changes. Namely, the sign of each ladder changes as compared to the Dirac magnons case, and, as a result, there is no way for the divergency to occur.
Moreover, away from $\Omega = 6SJ$, the pairing is only weakly suppressed by the rescattering processes.
However, when pump's frequency $\Omega$ approaches $6SJ$, $\alpha \rightarrow 0$, the pairing vanishes.

\subsubsection{Example: shifting the rescattered field away for $\Omega = 6SJ$}
When pump's frequency is $\Omega = 6SJ$ there is a resonant absorption of magnons. This can be see from $ {\cal L}^{\mathrm{R}/\mathrm{A}}_{\alpha\beta,0,\Omega}{\cal L}^{\mathrm{R}/\mathrm{A}}_{\beta\alpha,0,\Omega}   - {\cal L}^{\mathrm{R}/\mathrm{A}}_{\beta\beta,0,\Omega}{\cal L}^{\mathrm{R}/\mathrm{A}}_{\alpha\alpha,0,\Omega} = 0 \mp i0$ for non-interacting magnons. Upon inserting life-time of magnons at $\omega = 6SJ$ and ${\bf k} = 0$, the quantity becomes finite, imaginary and can be large. Let us call it 
\begin{align}
{\cal L}^{\mathrm{R}/\mathrm{A}}_{\alpha\beta,0,\Omega}{\cal L}^{\mathrm{R}/\mathrm{A}}_{\beta\alpha,0,\Omega}   - {\cal L}^{\mathrm{R}/\mathrm{A}}_{\beta\beta,0,\Omega}{\cal L}^{\mathrm{R}/\mathrm{A}}_{\alpha\alpha,0,\Omega} 
= \mp \frac{i}{2\tau_{6}}(6SJ \pm \frac{i}{2\tau_{6}}).
\end{align}
Also
\begin{align}
&
{\cal L}^{\mathrm{R}/\mathrm{A}}_{\beta\alpha,0,\Omega}-
{\cal L}^{\mathrm{R}/\mathrm{A}}_{\beta\beta,0,\Omega} = \mp \frac{i}{2\tau_{6}},
\\
&
{\cal L}^{\mathrm{R}/\mathrm{A}}_{\alpha\beta,0,\Omega}-
{\cal L}^{\mathrm{R}/\mathrm{A}}_{\alpha\alpha,0,\Omega} = \mp \frac{i}{2\tau_{6}},
\end{align}
and, hence, we get
\begin{align}
\frac{{\cal L}^{\mathrm{R}/\mathrm{A}}_{\beta\alpha,0,\Omega}-
{\cal L}^{\mathrm{R}/\mathrm{A}}_{\beta\beta,0,\Omega}}{{\cal L}^{\mathrm{R}/\mathrm{A}}_{\alpha\beta,0,\Omega}{\cal L}^{\mathrm{R}/\mathrm{A}}_{\beta\alpha,0,\Omega}   - {\cal L}^{\mathrm{R}/\mathrm{A}}_{\beta\beta,0,\Omega}{\cal L}^{\mathrm{R}/\mathrm{A}}_{\alpha\alpha,0,\Omega} } 
= \frac{1}{6SJ\pm \frac{i}{2\tau_{6}}}.
\end{align}
Therefore, the shift of the $\omega = 6SJ$, ${\bf k}=0$ fields reads as
\begin{align}
&
\bar{\Psi}^{\mathrm{cl}}_{n;0;6SJ} 
\rightarrow
\bar{\Psi}^{\mathrm{cl}}_{n;0;6SJ}  + \frac{\Gamma \sqrt{S}}{6SJ-\frac{i}{2\tau_{6}}},
\\
&
\Psi^{\mathrm{cl}}_{n;0;6SJ} 
\rightarrow
\Psi^{\mathrm{cl}}_{n;0;6SJ}  + \frac{\Gamma \sqrt{S}}{6SJ+\frac{i}{2\tau_{6}}}.
\end{align}
For physically relevant scenario, $6SJ>\frac{1}{2\tau_{6}}$, thus, we can neglect the inverse life-time, and recover the claim made in the Main Text.

\section{Supplemental Material References}
\begin{enumerate}
\item A.I. Akhiezer, V.G. Bar'yakhtar, and S.V. Peletminskii, \textit{Spin Waves} (Nauka, Moscow, in Russian, 1967).

\item A. Auerbach, \textit{Interacting Electrons and Quantum Magnetism} (Springer, New York, 1994).

\item S.M. Rezende, \textit{Fundamentals of magnonics} (Springer, 2020).

\item A. Kamenev, \textit{Field theory of non-equilibrium systems} (Cambridge, University Press, 2012).

\item H. Suhl, J. Phys. Chem. Solids, {\bf 1}, 209 (1957).

\item M. Bloch, Phys. Rev. Lett. {\bf 9}, 286 (1962)

\item S.S. Pershoguba, S. Banerjee, C. Lashley, J. Park, H. \AA gren, G. Aeppli, and A.V. Balatsky, Phys. Rev. X {\bf 8}, 011010 (2018).

\end{enumerate}

\end{widetext}
\end{document}